\documentclass[10pt,a4paper]{article}

\usepackage[T1]{fontenc}
\usepackage[affil-it]{authblk}

\usepackage{amsmath}
\usepackage{amsfonts}
\usepackage{multirow}
\usepackage{multicol}
\usepackage{graphicx}
\usepackage{subfigure}
\usepackage{url}
\usepackage{enumitem}

\usepackage[table]{xcolor}

\newcommand{\mk}[1]{{\color{magenta}#1}}

\usepackage{amsmath}
\usepackage{amsfonts}
\usepackage{multirow}
\usepackage{multicol}
\usepackage{graphicx}
\usepackage{subfigure}
\usepackage{wrapfig}
\usepackage{rotating}
\usepackage{times}
\usepackage{xspace}
\usepackage{amsbsy}
\usepackage{color}

\def\<{\langle}
\def\>{\rangle}

\def\be{\begin{equation}}
\def\ee{\end{equation}}
\def\bea{\begin{eqnarray}}
\def\eea{\end{eqnarray}}

\renewcommand{\eqref}[1]{Eq. (\ref{eq:#1})}

\makeatletter
\DeclareRobustCommand\onedot{\futurelet\@let@token\@onedot}
\def\@onedot{\ifx\@let@token.\else.\null\fi\xspace}

\def\eg{\emph{e.g}\onedot} 
\def\ie{\emph{i.e}\onedot} 
 
 \def\vs{\emph{vs}\onedot}

\newcounter{iictr}
\def\@iia[#1]{\setcounter{iictr}{#1}\@iib}
\def\@iib{\iii[\roman{iictr}]\xspace}
\def\ii{\@ifnextchar[{\@iia}{\stepcounter{iictr}\@iib}}
\def\iii[#1]{({\em#1\/})}
\makeatother

\title{Unsupervised Visual and Textual Information Fusion 
in Multimedia Retrieval - A Graph-based Point of View\footnote{An extended version of the paper 
Visual and Textual Information Fusion in Multimedia Retrieval using Semantic
  Filtering and Graph based Methods, by J. Ah-Pine, G. Csurka and S. Clinchant,
  submitted to ACM Transactions on Information Systems.}}

\author[1]{Gabriela Csurka}
\author[2]{Julien Ah-Pine}
\author[1]{St\'ephane Clinchant}

\affil[1]{Xerox Research Centre Europe, 6 chemin de Maupertuis \\
38240,  Meylan France \\
\texttt{Firstname.Lastname@xrce.xerox.com}
}

\affil[2]{University of Lyon 2, ERIC Lab, 5, avenue Pierre Mendes France \\
69676 Bron Cedex, France\\
\texttt{julien.ah-pine@eric.univ-lyon2.fr} }

\date{}

\begin{document}


\maketitle

\begin{abstract}  

Multimedia collections are more than ever growing in size and diversity. Effective multimedia retrieval systems are thus critical to access these datasets from the end-user perspective and in a scalable way. We are interested in repositories of image/text multimedia objects and we study multimodal information fusion techniques in the context of content based multimedia information retrieval. We focus on graph based methods which have proven to provide state-of-the-art performances. We particularly examine two of such methods : cross-media similarities and random walk based scores. From a theoretical viewpoint, we propose a unifying graph based framework which encompasses the two aforementioned approaches. Our proposal allows us to highlight the core features one should consider when using a graph based technique for the combination of visual and textual information. We compare cross-media and random walk based results using three different real-world datasets. From a practical standpoint, our extended empirical analysis allow us to provide insights and guidelines about the use of graph based methods for multimodal information fusion in content based multimedia information retrieval.  
\end{abstract}





\maketitle

\section{Introduction}

With the continuous growth of communication technologies, the information that
we consult, produce and communicate whatever the communication device we use,
has been richer and richer in terms of the media it is composed of. The web has
particularly contributed to the production of such multimedia or multimodal
data. For instance, web pages from news agencies websites are texts illustrated with pictures or videos; photo sharing websites, such as FlickR, store pictures annotated with tags; video hosting websites, such as Youtube, are again examples of multimedia data repositories. Apart from the web, we have also witnessed the development of new services that rely on digital libraries made of data composed of several media. In museums for example, there are more and more multimedia applications using text, image, video and speech in order to better plunge the visitor into the historical context of the piece of art she is consulting. New generations of television devices now propose on-line interactive media, 
on-demand streaming media and so on. The ever-growing production of multimodal data has brought the multimedia research community to address the problem of effectively accessing multimedia repositories from the end-user perspective and in a scalable way. Accordingly, multimedia data search has been a very active research domain for the last decades.

There are different ways to search a multimedia repository. As for videos or
images datasets such as Youtube or FlickR, we typically index those media by
means of the title, metadata, tags or text associated to or surrounding
them. Then, we search those multimodal objects by using text queries and text
based search engines. There are different reasons we use text to retrieve videos
or images. Firstly, it is not always possible for the user to query  a
collection by examples, since the search engine cannot always provide her with
videos or images that represent the type of items she would like to
retrieve. Secondly, videos or images are stored in machines into a computational
representation consisting of low-level features which do not carry by their own
the high level semantics.  In other words, it is a strong challenge to
effectively associate low-level features extracted from videos or images with
high-level features such as keywords or tags without using pre-trained
classifiers.  This problem is known as the semantic gap. As a consequence of those two difficulties, we generally use the text media for content based multimedia information retrieval (CBMIR) in order to have more relevant search results.

If the text is the core media to use in order to access a multimedia repository effectively, it is however beneficial to use other media in addition to the text, during the course of the search process. Indeed, most of research works about multimedia information fusion have shown that combining different modalities to address CBMIR tasks, even with simple strategies, is beneficial. In this paper we are interested in this topic. We particularly address the combination of visual and textual information. We thus deal with repositories that are composed of multimedia objects made of an image associated with a text. There are different multimedia information fusion methods and in this paper we are interested in graph based techniques. Such approaches became very popular in the information retrieval community since the development of techniques like PageRank or Hits \cite{Brin:1998:ALH:297810.297827,Kleinberg:1999:ASH:324133.324140,Langville:2005:SEM:1055334.1055396}.

In a nutshell, the goals of this paper are the following ones :
\begin{itemize}[label=\textbullet]
\item We discuss the semantic filtering method that seeks to enhance the similarities between multimedia items when they are composed of both a visual and a textual part \cite{Clinchant_etal_icmr11}. We explain how such a filter based on the text query can better cope with the semantic gap in the case of CBMIR. We propose to use this approach as a first level of the fusion process of visual and textual information in our multimedia relevance model. Indeed, not only the proposed semantic filtering improves the similarity measures between multimedia items but it also allows reducing the storage and computational complexities of graph based models.
 \item We study and compare two popular graph based multimedia information fusion methods that were originally proposed in two different research communities. On the one hand, we analyze the cross-media similarity approach initially proposed in \cite{Clinchant_wn07,Clinchant_pr07} for content based image retrieval in the context of ImageCLEF multimedia retrieval tasks. On the other hand, we investigate the random walk approach which was initially proposed in \cite{DBLP:conf/mm/HsuKC07,DBLP:journals/ieeemm/HsuKC07} and used on several TRECVID tasks for content based video retrieval. Our main contribution in that perspective is to show that the two techniques are related. In fact, we propose a unifying framework that generalizes both approaches. This generalization allows us to better compare the two techniques, to propose a third approach which amounts to a mix of both latter methods, and it also aims at examining the main points and settings when using graph based methods for the combination of visual and textual information in CBMIR.
 \item We analyze two different multimodal search scenarios. In the first
 scenario, we suppose that the user can only use a text query in order to
 retrieve images. Multimedia objects of the repository are indexed using their
 text part and a text based search engine is used in a first time. In a second
 time, we use the visual information of the (text based) retrieved objects in order to improve the
 search results. This multimedia search scenario is referred along this paper as
 {\bf the asymmetric case} since the user can only use a text query. In contrast, in
 the second scenario, the user can use a multimedia query which means that she
 can enter a text query accompanied with one or several images as examples of
 her information need. To this second scenario we refer as {\bf the symmetric search scenario}. 
 \item We experiment with 3 different image/text datasets which have distinct features. We conduct many tests in order to have a better analysis of the core points in the use of the graph based methods under study and in the context of image/text multimedia retrieval. Our experimental results allow us to provide insights and guidelines about how to set the parameters of the unified graph based technique we propose.
\end{itemize}

The rest of this paper is organized as follows. In section \ref{general_background} we review the main families of multimodal fusion approaches and their features. We take into consideration both the asymmetric and the symmetric search scenarios. In section \ref{graph_based}, we discuss the use of graph based methods to fuse visual and textual information and we detail the cross-media similarities and the random walk based techniques which are the techniques under study in this paper. Next, in section \ref{sec:sem_filt}, we introduce the semantic filtering method which represents a core step to refine multimedia similarities from a semantic standpoint, when textual information is available. Such an approach amounts to a first level of multimedia information fusion and in addition, it also enables reducing the storage and computational complexities of graph based methods. In section \ref{comp_cm_rw}, in light of the material exposed in sections \ref{graph_based} and \ref{sec:sem_filt}, we introduce our multimedia relevance model that relies both on the semantic filtering guided by the text query and an unifying graph based framework that embodies both cross-media similarities and random walks based scores.
Then, we describe in section \ref{subsec.exp_settings}, the experimental settings we conducted on three real-world multimedia collections in order to validate our work. We introduce the image and text representations and the monomedia similarity matrices we used. In section \ref{experiments}, we  present the experimental results we obtained with the different tested fusion strategies in the goal of comparing cross-media similarities and random walk based scores. We finally discuss some other advantages of our proposal in terms of complexity as regard to large collections and we provide some guidelines on how to use the generalized graph based approach we propose. In section \ref{discussions}, we summarize our main findings.

\section{Families of unsupervised multimedia fusion techniques in CBMIR}\label{general_background}

A good introduction to the multimedia information access domain, its challenges
and its basic techniques can be found in \cite{Rueger_book} that covers the common topics in multimedia IR such as feature extraction, distance measures, supervised classification also known as automatic tagging and fusion of different experts.
In this paper, we are particularly interested in multimedia fusion techniques and the literature on this topic is very vast. In this section we attempt to depict the main families of fusion methods for visual and textual information. It is important to precise that we place ourselves in an unsupervised context meaning that we do not use any learning technique in our framework. We can mention at least two research communities that have been addressing this research topic actively. On the one hand, there are the research teams that have participated in the TRECVID workshop series and have focused their research efforts on video retrieval \cite{1178722}. On the other hand, we can quote the research groups involved in the ImageCLEF meetings and which have been interested in the tracks related to image and multimedia retrieval \cite{ImageClefBook10}. In the former research community it is usually assumed that the user does not have any example query and the common way to search a multimedia collection rely solely on textual queries. On the contrary, in the latter research community, it is generally assumed that the user information need is expressed by a multimedia query composed of an image query and a related text query. We present in the following, broad families of multimedia fusion techniques that have been studied for the two distinct search scenarios.

Despite the fact that we focus on unsupervised multimedia fusion, we also point to some research papers that address multimedia fusion techniques from a supervised or a semi-supervised perspective and which show some connections with our work.

\subsection{The symmetric case with an image query and a text query}\label{subsec.sym_scenario}

\begin{figure}[t]
\begin{center}
\includegraphics[width=11cm]{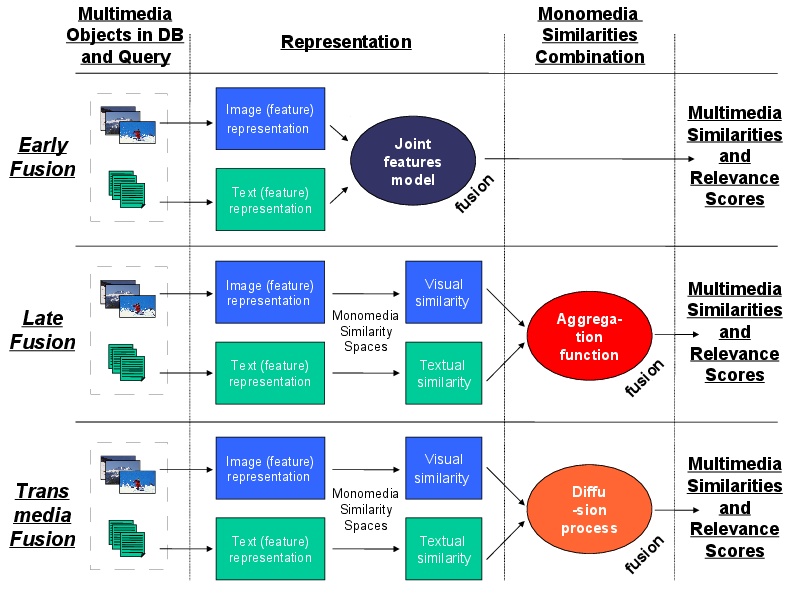}
\caption{Early, late and transmedia fusion.\label{fig:fusion}}
\end{center}
\end{figure}

Most of the techniques developed in this context fall in three different categories : early, late and transmedia fusion. We depict these three families of approaches by distinguishing the inherent steps they are composed of. This is summarized in Figure \ref{fig:fusion}. In the following, we assume that the multimedia query can be considered similar as any item of the multimedia collection that is to say an object made of an image part and a text part. Given a multimedia query, the search process consists in measuring a multimedia similarity between the query and the multimedia items in the repository.

The early fusion approach represents the multimedia objects in a multimodal feature space designed \textit{via} a joint model that attempts to map image based features to text based features and \textit{vice versa}. The simplest early fusion method consists in concatenating both image and text feature representations (see \eg \cite{SnoekICM2005,Clinchant07,ImageClefBook10}). However, more elaborated joint models such as Canonical Correlation Analysis have been investigated \cite{Mori99,Lavrenko03,Vinokourov03,RasiwasiaPCDLLV10}.
In the same vein, \cite{InfoTheo_rueger} presents an information theoretic framework that could  also fit into this family of fusion approaches. 

On the contrary, late fusion and transmedia fusion strategies do not act at the level of the monomedia feature representations but rather at the level of the monomedia similarities \cite{Clinchant_wn07,BrunoMM08}. In these contexts, we assume that we have effective monomedia similarities and that it is better to combine their respective decisions rather than attempting to bridge the semantic gap at the level of the features.

Concerning late fusion techniques, they mainly seek to merge the monomedia relevance scores by means of aggregation functions. In that case, the simplest aggregation technique used is the mean average \cite{EscalanteHSM08} but more elaborated approaches have been studied (\eg \cite{CaicedoMNG10,ImageClefBook10,CsurkaClinchant12,Wilkins_2010}). 

As far as transmedia fusion methods are concerned, they act like similarity diffusion processes. The resulting combination is non linear unlike most of late fusion techniques. These methods generally amount to mixing monomedia similarity matrices by means of matrix multiplication operations \cite{DBLP:conf/mm/WangMXL04,DBLP:conf/kdd/PanYFD04,DBLP:conf/mm/HsuKC07,Clinchant_wn07}. In that kind of relevance models, we usually start the diffusion process using pseudo-relevant items only. In that case, we typically use the $k$ nearest neighbors according to one monomedia relevance score and thus those methods are also inspired from the pseudo-relevance feedback mechanism in information retrieval (see \eg \cite{ruthven2003survey}). 

It is important to mention that there are other ways to categorize the different multimedia fusion techniques. In the recent survey paper \cite{DBLP:books/sp/mining2012/ZhaWSC12} for example, other terms are used. Nevertheless, they basically correspond to the definitions given above with the following mappings : early, late and transmedia fusion are named latent space based, linear fusion and graph based fusion in \cite{DBLP:books/sp/mining2012/ZhaWSC12}.

\subsection{The asymmetric case with a text query only}\label{subsec.vis_rerank}

In addition to the three previously recalled types of fusion methods, \cite{DBLP:books/sp/mining2012/ZhaWSC12} cites another category named visual reranking. This fourth family of techniques assumes that multimedia collections are accessed using textual queries solely. Therefore, in this context, there is an explicit asymmetry between image and text in the multimedia search scenario.

\begin{figure}[t]
\begin{center}
\includegraphics[width=11cm]{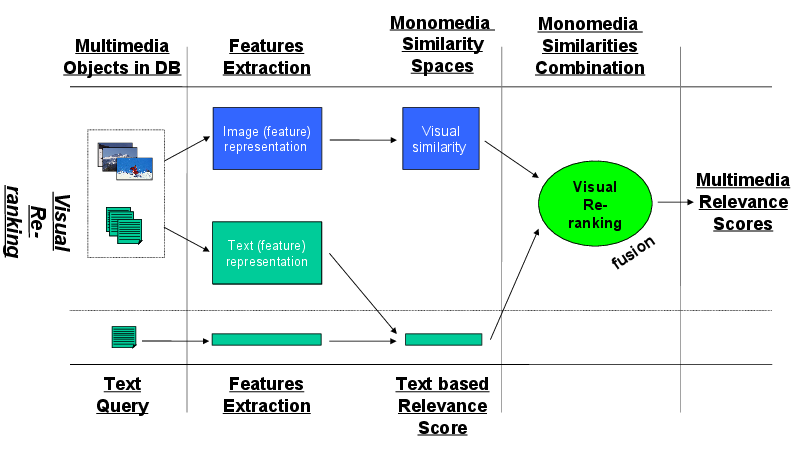}
\caption{Visual reranking.\label{fig:vis_rerank}}
\end{center}
\end{figure}

Visual reranking techniques particularly deals with such a search scenario. They proceed in two steps : using the text query, they first use text based similarities in order to find the most relevant objects from a semantic viewpoint; then, they employ the visual similarities between objects of the database in order to refine the textual similarities based ranking. Similarly to Figure \ref{fig:fusion}, we depict in Figure \ref{fig:vis_rerank} the different main steps of visual reranking approaches.

The common assumption that all visual reranking techniques make is that visually similar images should have similar relevance scores \cite{DBLP:conf/mm/MoriokaW11}. However, different approaches are used to re-arrange the top retrieved items by the text similarities in order to take this principle into account. According to \cite{DBLP:books/sp/mining2012/ZhaWSC12}, we can categorize visual reranking techniques into three subcategories : classification based, clustering based and graph based.

In the first case, pseudo-positive and pseudo-negative objects are sampled from the text based ranked list then a learning to rank algorithm is trained on the visual features (see \eg \cite{Liu:2009:LRI:1618303.1618304} for a general reference on learning to rank methods). Afterward, objects are re-ordered according to the scores provided by the trained classifier. The critical point is the sampling method used to select pseudo-training examples. The simplest strategy considers items at the top of the list as pseudo-positive and items at the bottom as pseudo-negative but more sophisticated approaches have been proposed \cite{DBLP:conf/mm/TianYWYWH08,DBLP:conf/mm/YangH10,DBLP:conf/mm/MoriokaW11}.

As for clustering based visual reranking, the main idea is to cluster the list of text based retrieved items and to re-arrange them such that objects that are visually highly similar and have high initial text retrieval scores are favored \cite{DBLP:conf/mm/HsuKC06,DBLP:journals/ieeemm/HsuKC07}.

Graph based methods consider multimedia objects as nodes of a graph and the
different types of relationships they share as edges. Examples of weighted edges
between objects are visual similarities or textual similarities but depending on
the application other types of relations can be considered. Graph analysis
techniques are then employed in order to infer new features in the goal of
re-arranging the text based ranked list of items. One such method, inspired
by the well-known PageRank
\cite{BrinPage98,Langville:2005:SEM:1055334.1055396,Franceschet11}  
used to rerank web pages by search engines such as Google, was 
proposed in \cite{DBLP:conf/mm/HsuKC07,DBLP:journals/ieeemm/HsuKC07}. 
It is based on random walks over a stochastic matrix which is deduced 
from the fusion of visual and textual similarities, and  the stationary 
probability distribution over the nodes is then additionally used 
to rerank the initial retrieved list.
In the same vein, \cite{Craswell2007} proposed a 
Markov random walk model with backward and forward steps. 
They found out that the best performances were obtained
with a long backward walk with high self-transition probability.

\subsection{Graph based techniques in both search scenarios}\label{subsec.graphbased}

Transmedia fusion techniques we introduced in paragraph
\ref{subsec.sym_scenario} are technically similar to graph based methods
presented in the previous paragraph. Indeed, both approaches use similarity
matrices to respectively rank or rerank multimedia items. Graph based methods
have proven to be state-of-the-art techniques for many information retrieval
tasks (see \eg \cite{BrinPage98,Langville:2005:SEM:1055334.1055396,Franceschet11}). In CBMIR too, they have demonstrated their advantages over early or late fusion approaches in many research works (see \eg \cite{ImageClefBook10,DBLP:books/sp/mining2012/ZhaWSC12}). We thus focus on such methods in this paper. 

Besides, there has been very few research works that address CBMIR in a symmetric search scenario and using graph based methods. Consequently, in this paper we study the different kinds of search scenarios with such techniques in order to have a better comparison between the asymmetric and the symmetric search scenarios in this context. 

Before presenting in more details the two graph based fusion techniques we examine in the rest of the paper, we present in the next paragraph some additional references that also tackle multimedia information fusion and/or multimedia retrieval but in other learning settings.

\subsection{Multimedia fusion in a supervised or a semi-supervised context}\label{subsec.supervised}

We review some related research papers that tackle video and image search from a
multimodal perspective but employing supervised or semi-supervised
techniques. In \cite{6212356} for example, the authors use hypergraph learning
to design a joint visual-textual representation of multimedia objects. This
method amounts to an early fusion scheme. Another early fusion approach was
presented in \cite{Natsev:2007:SCQ:1291233.1291448} and which addresses
multimedia query expansion for both the text and the image parts. This work
relies on an intermediate representation of multimedia information in a
predefined visual-concept lexicon. Classification models are used to map the
queries to the lexicon.  Then based on  pseudo-relevance feedback 
different query expansion  and score reranking methods are proposed. 
Similarly, \cite{TorresaniFitzgibbonSigir13} uses intermediate 
representation, in their case  visual classifiers. To build these classifiers
they download images from Google  or Bing using query words and 
represent these images by classemes (attribute-based image descriptors). 
In a second step, images  in the  web pages are classified using these
classifiers and the scores are used to rerank the multimodal
documents (in their cases the web pages). The reranking is also supervised as 
a  set of  training queries with relevance scores are used 
to learn the parameters of the latter algorithm.

Other related works that are worth mentioning are the following ones \cite{4785118,4801611,6242410}. These papers address video semantic annotation and web image search in a semi-supervised fashion. The general framework used in these contributions is formulated as an optimization problem that simultaneously deal with the late fusion of monomedia similarity matrices and graph semi-supervised learning. The solutions of the optimization problems can be formulated using normalized graphs Laplacian and iterated algorithms are proposed to infer the relevance scores which are further used for annotating videos or ranking images. 

The main differences between these research works and our framework are the
following ones~: (i) we do not use any learning models nor external resources
(such as a domain ontology or downloaded image set) and we only rely on the surrounding text of images which is a more general setting; (ii) we emphasize the transmedia principle in the diffusion process which mix the monomedia similarity matrices and relevance scores differently from late fusion; (iii) since no learning phase is required in our case we avoid the annotation burden and also the time complexity problem underlying such methods.

After having introduced a classification of the most used unsupervised multimedia information fusion strategies and discussed some other related works, we introduce in the next section, the graph based fusion methods we are going to embed in our multimedia relevance model.

\section{Cross-media similarities and random walk based scores}\label{graph_based}

We recall two popular image/text graph based fusion techniques in CBMIR and we consider their use in the two different search scenarios we recall previously. The first approach called cross-media similarities was proposed in the context of ImageCLEF workshop series while the second method based on random walks and called context reranking was used in TRECVID tasks. 

For convenience, we introduce in Table \ref{tab:notations} the notations we will use in the rest of the paper. Note that we assume that the different similarities or scores are all non negative numbers.

\begin{table}[t]
\begin{center}
\begin{tabular}{|r|l|}
\hline
Notations & Definitions \\
\hline
$v$ & Subscript indicating the visual part of an entity \\
$t$ & Subscript indicating the textual part of an entity\\
$q=(q_v,q_t)$ & Multimedia query (which reduces to $q=(q_t)$ in the asymmetric search scenario)\\
$d=(d_v,d_t)$ & Multimedia object in the database\\
$n$ & The number of multimedia objects or documents in the database\\
$s_v(q,.)$ & Visual similarities (row) vector of $q$ with all documents of the database (of size $1\times n$)\\
$s_t(q,.)$ & Textual similarities (row) vector of $q$  (of size $1\times n$)\\
$l$ & The number of top elements retained from $s_t$ for semantic filtering. \\
$S_v$ & Visual similarity (square) matrix between pairs of documents (of size $n\times n$)\\
$S_t$ & Textual similarity (square) matrix between pairs of documents (of size
$n\times n$)\\
$s^{q_t}_*, S^{q_t}_*$ & Same as above but text query semantically filtered 
(of size $1\times l$ and $l\times l$)\\
$\mathbf{K}(.,k)$ & $k$ nearest neighbor thresholding operator acting on a
vector \\ 
$x_{(i)}$ & Diffusion process iteration on the full graph,
starting from the text modality (of size $1\times n$)\\
$y_{(i)}$ & Diffusion process iteration on the full graph,
starting from the visual modality (of size $1\times n$)\\
$x^{q_t}_{(i)}, y^{q_t}_{(i)}$ & Same as above but using the graph reduced with
the $q_t$ based semantic filter (of size $1\times l$)\\
$cm^{q_t}_{tv}, cm^{q_t}_{vt}$ & Cross-media similarities corresponding to
$x^{q_t}_{(1)}$ respectively to  $y^{q_t}_{(1)}$ \\
$rw^{q_t}_{tv},rw^{q_t}_{vt}$ & Random walk based scores corresponding to
$x^{q_t}_{(\infty)}$ respectively to  $y^{q_t}_{(\infty)}$, with $k=l$ \\
$gd^{q_t}_{tv},gd^{q_t}_{vt}$ & Generalized diffusion model corresponding to
 $x^{q_t}_{(\infty)}$ respectively to  $y^{q_t}_{(\infty)}$, 
with $k \ll l$.\\
\hline
\end{tabular}
\caption{Notations and definitions.\label{tab:notations}}
\end{center}
\end{table}

\begin{figure}
\begin{center}
\includegraphics[width=0.65\textwidth]{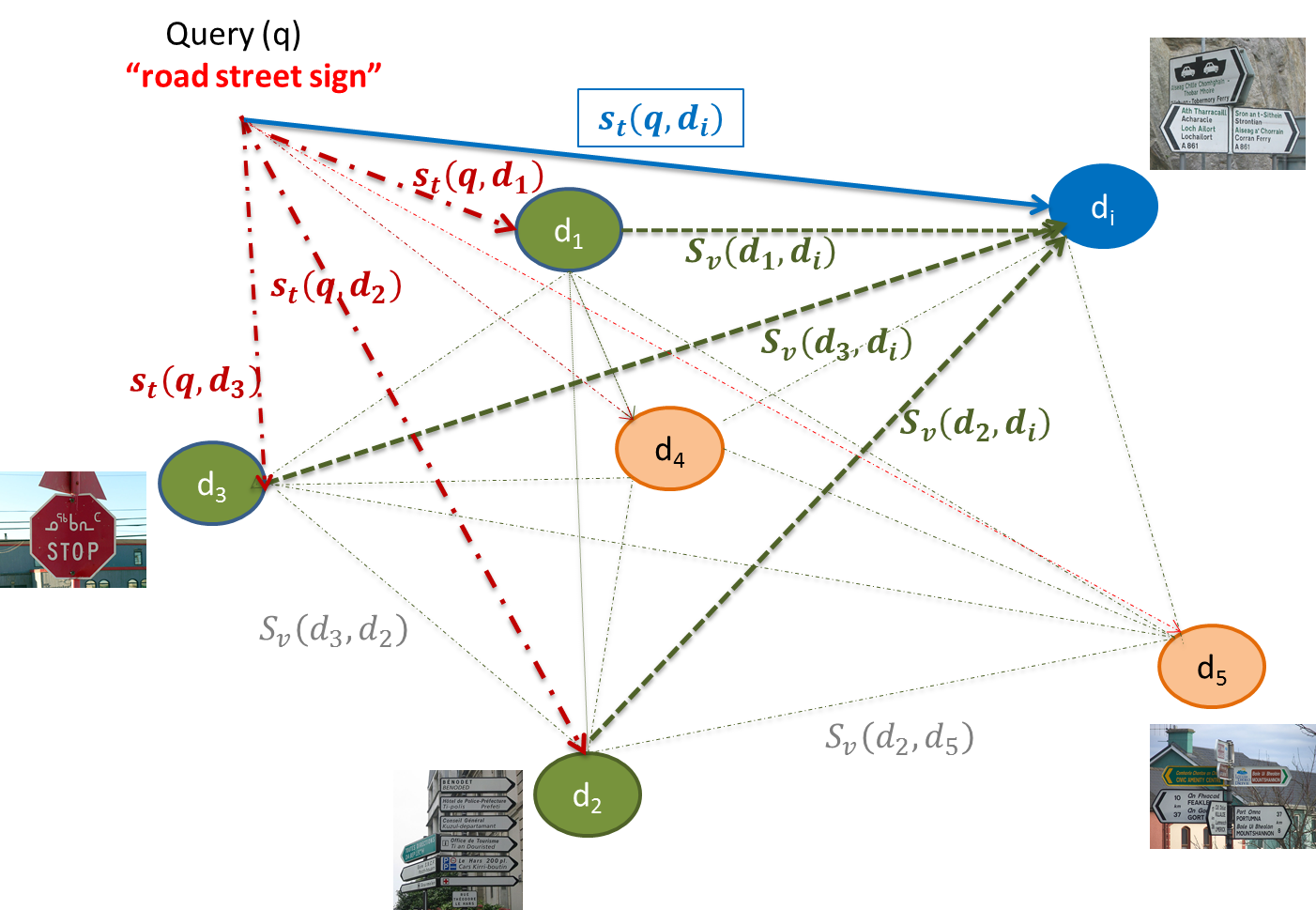}
\end{center}
\caption{Given a text query $q_t$ the cross-media relevance score 
can be computed as
$\sum_{d_j\in \mathcal{N}_t(q)}  s_t(q,d_j) \cdot  S_v(d_j,d_i)$. Note that the sum is over the nearest neighbors of 
the ``query'', hence the complementary visual information of the documents 
that are close to the query are exploited.}
\label{fig:CMgraph}
\end{figure}

\subsection{Methods based on cross-media similarities}\label{cross_media}

Cross-media similarities studied in this paper refer to the research work developed in the following references \cite{Clinchant_wn07,ah_al_mta09} and which has proven to give top-ranked retrieval results on several ImageCLEF multimedia search tasks\footnote{For more details, please visit \url{www.imageclef.org}} \cite{ImageClefBook10}.

We can explain the cross-media similarity mechanism using the following
illustration (see also  Figure \ref{fig:CMgraph}). Given a text query $q_t$, we first find the most similar items in the collection with regard to the textual similarities. Then, we select pseudo-relevant objects $d$ which are the set of $k$ nearest neighbors. Next, we look at the pseudo-relevant objects' visual similarities profiles $S_v(d,.)$. We then combine these visual similarity scores linearly and we obtain a cross-media similarity measure between the text query and the multimedia objects of the database. Formally such cross-media similarities are defined as follows~:
\begin{eqnarray}\label{eq.ctv}
cm_{tv}(q,.)=\mathbf{K}(s_t(q,.),k)\cdot S_v
\end{eqnarray}
where~:
\begin{itemize}
\newcommand{\myitem}{\item[\textbullet]}
 \myitem $\mathbf{K}(.,k)$ is an operator that takes as input a vector 
and  gives a zero value to elements whose score is strictly lower then the
 $k^{th}$ highest score. 
 \myitem The $\cdot$ symbol represents the regular matrix multiplication operation.
\end{itemize}

The previously introduced cross-media similarity, denoted $cm_{tv}(q,.)$, propagates the text similarities of pseudo-relevant objects to their visual similarities which can be seen as a transmedia pseudo-relevance feedback mechanism. This operation is non commutative and we can design a cross-media similarity, $cm_{vt}(q,.)$, propagating visual similarities to textual similarities, providing that we are also given an image query $q_v$. We then obtain~:
\begin{eqnarray}\label{eq.cvt}
cm_{vt}(q,.)=\mathbf{K}(s_v(q,.),k)\cdot S_t
\end{eqnarray}

These cross-media similarities attempt to bridge the semantic gap between visual and textual information by enriching one modality by the other using monomedia nearest neighbors as proxies. 
Once the cross-media similarities are computed we can linearly combine them with monomedia similarities as follows~:
\begin{eqnarray}\label{cm_formula}
rsv_{cm}(q,.)= \alpha_t s_t(q,.)+\alpha_v s_v(q,.)+\alpha_{tv} cm_{tv}(q,.)+ \alpha_{vt} cm_{vt}(q,.)
\end{eqnarray}
where $\alpha_t,\alpha_v,\alpha_{tv},\alpha_{vt}$ are real parameters that sum to one.

The formula given in Eq. \ref{cm_formula} encompasses different particular sub-cases~:
\begin{itemize}
\newcommand{\myitem}{\item[\textbullet]}
 \myitem $\alpha_{tv}=\alpha_{vt}=0$, leads to the classic late fusion technique using a weighted mean as an aggregation function.
 \myitem $\alpha_{v}=\alpha_{vt}=0$, gives a cross-media based approach to address CBMIR tasks in the context of the asymmetric case.
 \myitem $\alpha_{v}=\alpha_{tv}=0$, is one particular combination that gave top-ranked results on different ImageCLEF tasks \cite{Clinchant_wn07,ap_al_wn08,ap_al_wn09}. Indeed, it was already shown that the visual information is particularly beneficial through the cross-media $cm_{vt}$ scores that spread visual information to textual information.
\end{itemize} 

Cross-media similarities draw inspiration from Cross-Media Relevance
Models \cite{Jeon_crossmedia_rm} and intermedia feedback methods proposed in \cite{chevallet_interfeedback}.
For example, from an image query, a first visual similarity is computed and an initial set of (assumed)
relevant objects is retrieved. As the objects are multimodal, each image has also
a text part, and this text can feed any text feedback method (other than relevance
models). In other words, the modality of data is switched, 
from image to text or text to image, during the (pseudo) feedback process. 
In that sense, cross-media techniques generalize the pseudo-feedback 
idea present in the cross-media relevance model.

\subsection{Methods based on random walks}\label{random_walk}

\begin{figure}
\begin{center}
\includegraphics[width=0.65\textwidth]{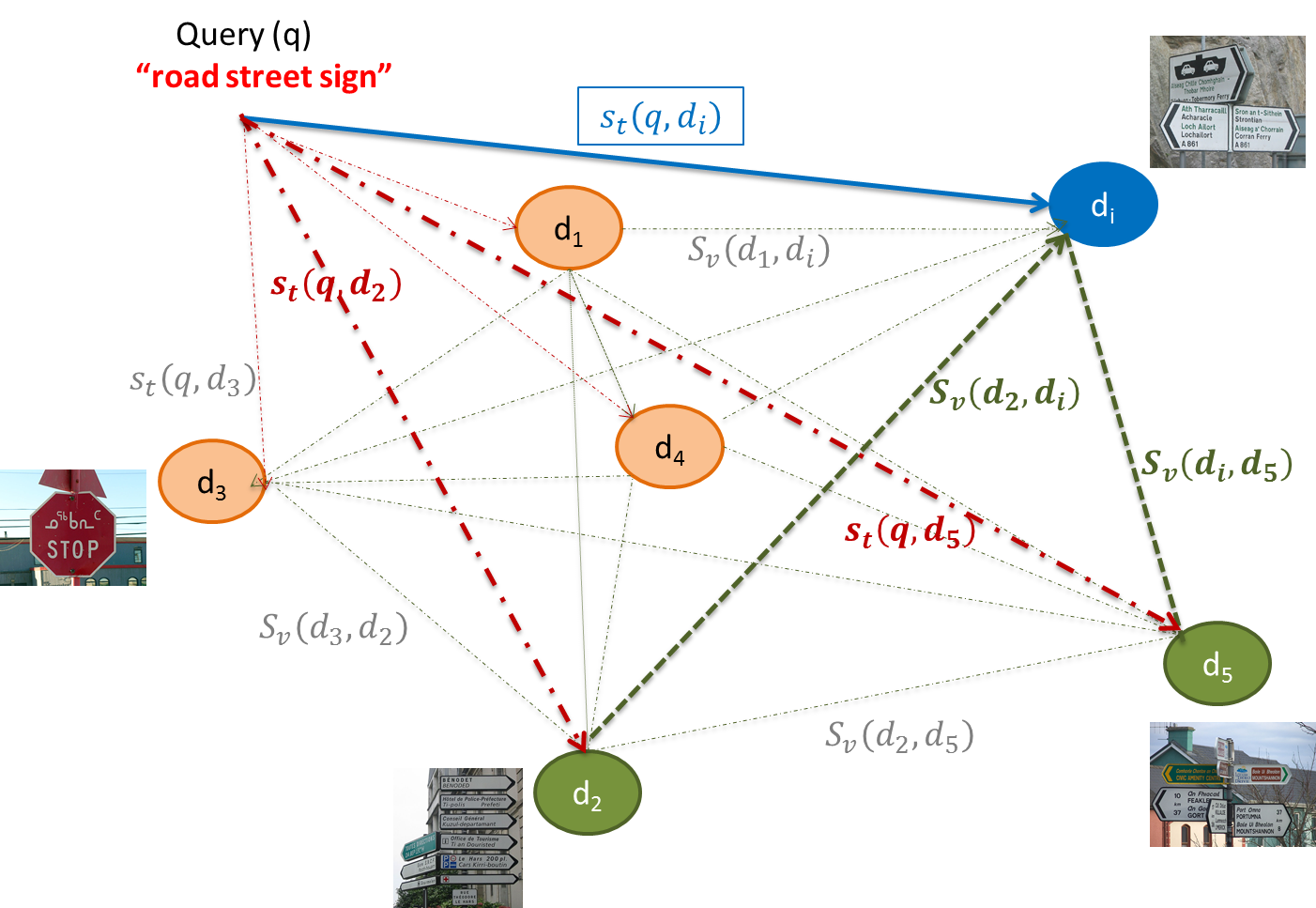}
\end{center}
\caption{Given a text query $q_t$ the new relevance score can be computed as
$s_t(q,d_i)+ s_t(q,d_i) + \sum_{d_j\in \mathcal{N}_v(d_i)}  
s_t(q,d_j) \cdot S_v(d_j,d_i)$. Note that in contrast 
to the cross-media, the sum is over the  nearest neighbors of  the 
document $d_i$, hence the visual (or multi-modal if 
$P=(1-\beta) S_v +\beta S_t$ are  considered)  context of the  document 
$d_i$ is exploited.}
\label{fig:RWgraph}
\end{figure}

The PageRank algorithm proposed in 
\cite{BrinPage98,Langville:2005:SEM:1055334.1055396,Franceschet11} has been 
an important step forward in development and success of search 
engines such as Google.  It is therefore not surprising that 
multimedia information fusion 
based on graph modeling  using random walks  has
been addressed by several researchers
\cite{DBLP:conf/kdd/PanYFD04,DBLP:conf/mm/HsuKC07,DBLP:conf/mm/TianYWYWH08,DBLP:journals/tmm/MaZLK10}.
In this paper we particularly study the method proposed in
\cite{DBLP:conf/mm/HsuKC07,DBLP:journals/ieeemm/HsuKC07}. In this approach, it
is assumed that each image is a node of a graph and two images are linked with a
weighted edge if there exists a multimodal contextual similarity between them
((see also Figure \ref{fig:RWgraph}). Depending on the application, 
the definition of such multimodal contextual similarities can vary. 
Typically, we assume that they are given by a linear combination of some visual and textual similarities.

The research work described in \cite{DBLP:conf/mm/HsuKC07} deals with video retrieval. In the latter paper, the authors propose to use near-duplicate detection measures as for visual similarities between video stories. Text similarities are derived from automatic speech recognition and machine translation transcripts and measured by a mutual information approach.

In our perspective, we are concerned with image/text data and we assume generic
image based and text based similarity matrices which are respectively denoted
$S_v$ and $S_t$. Using the notations given in Table \ref{tab:notations}, the multimodal contextual similarity matrix according to \cite{DBLP:conf/mm/HsuKC07}, that we denote by $C$, can be interpreted as follows~:
\begin{eqnarray}\label{eq.Hsu}
 C=(1-\beta) S_v+ \beta S_t
\end{eqnarray}
where $\beta\in [0,1]$.

We then transform $C$ into a stochastic matrix, denoted by $P$, by applying the
following normalization operator\footnote{Note that before normalization, 
$P$ can be sparsified, \ie  only top pairwise similarities are
considered for each document as shown in Figure \ref{fig:RWgraph}. 
When $P$ is fully computed, it means that 
the context of each document is the whole dataset.}~:
\begin{eqnarray}\label{eq.sthnorm} 
 P= D \cdot C
\end{eqnarray}
where $D$ is the diagonal matrix of size $n\times n$, with general term
$D(i,i)=1/\sum_{j=1}^nC(i,j)$ and $D(i,j)=0$ for all $i\neq j$.

The general term $P(i,j)$ is interpreted as the probability to go from ``state''
$i$ to ``state'' $j$ where these indices respectively refer to documents $d^i$
and $d^j$. We then compute the random walk's stationary probability distribution
over the documents. Such graph based measures are then employed to rerank the
list retrieved by the text based scores. However, to further fuse visual and
textual information, the random walk is biased towards documents with higher
textual similarity values with the text query. In other words, we add a prior
based on the text scores in the random walk process. Note that such a prior can
also be interpreted as a restart process or a personalization vector in other
information retrieval tasks.

Formally, if we denote by $x_{(i)}$ the row vector of size $1\times n$ of the state probabilities at iteration $i$ then we have~:
\begin{eqnarray}\label{eq.it_rw_tv}
 x_{(i)}=(1-\gamma) x_{(i-1)}\cdot P + \gamma s_t(q,.)
\end{eqnarray}
where $\gamma\in [0,1]$.

In order to obtain the state stationary distribution, we iterate the previous updating equation until convergence which yields to the following definition~:
\begin{eqnarray}\label{eq.rw_tv}
 x_{\infty}=(1-\gamma) x_{\infty}\cdot P + \gamma s_t(q,.)
\end{eqnarray}

In \cite{DBLP:conf/mm/HsuKC07} only the asymmetric search scenario with a text query solely was treated. In this paper, we consider the extension of this approach when we are also given an image query. Accordingly, we use a similar random walk process but with a prior depending on the initial image based scores $s_v(q,.)$ and define the related stationary distribution~:
\begin{eqnarray}\label{eq.rw_vt}
 y_{\infty}=(1-\gamma) y_{\infty}\cdot P + \gamma s_v(q,.)
\end{eqnarray}

Let us denote $rw_{tv}(q,.)=x_{\infty}$ and $rw_{vt}(q,.)=y_{\infty}$. We can linearly combine these graph based scores with the initial monomedia similarities and design the following final relevance score~:
\begin{eqnarray}\label{rw_formula}
rsv_{rw}(q,.)=\alpha_t s_t(q,.)+\alpha_v s_v(q,.)+\alpha_{tv} rw_{tv}(q,.)+\alpha_{vt} rw_{vt}(q,.)
\end{eqnarray}

We can consider the following particular cases~:
\begin{itemize}
\newcommand{\myitem}{\item[\textbullet]}
 \myitem $\alpha_{vt}=\alpha_{tv}=0$, leads to the classic late fusion technique as for cross-media similarities.
 \myitem $\alpha_{t}=\alpha_{v}=\alpha_{vt}=0$, is a combination that reduces to $rw_{tv}$. It assumes the asymmetric search scenario and was tested\footnote{This combination was named FRTP in \cite{DBLP:conf/mm/HsuKC07}} in \cite{DBLP:conf/mm/HsuKC07}.
 \myitem $\alpha_{v},\alpha_{vt}>0$, is, to our knowledge,  a new extension of the method which assumes the symmetric search scenario.
\end{itemize}

Before analyzing further the two graph methods we have introduced, we discuss in the sequel, an important aspect of the combination of visual and textual information in CBMIR. Our multimedia retrieval model, we are going to introduce in section \ref{comp_cm_rw}, results from the materials described both in the present and the next sections.

\section{Text query based semantic filtering of multimedia similarities}\label{sec:sem_filt}

We first underline the particular importance of textual similarities between the
text query $q_t$ and the text part of multimedia items of the database when
addressing CBMIR tasks. Our observations lead us to propose the text based
semantic filtering of multimedia similarities that we argue to be a crucial
pre-processing step in CBMIR and thus in our multimedia retrieval model. As we
shall see, this approach is similar in spirit to the visual reranking
paradigm. However, in our perspective, we generalize the latter concept by
applying yhe semantic filtering not only to the visual similarities between the
query and the documents in the collection
but also to any  similarities we employ in our fusion model, such as visual or
textual similarities between the documents in the collection.
Before formally stating the text query based semantic filtering method, we
provide the rationale of such an approach by discussing the semantics 
conveyed by textual similarities as compared to visual similarities.

When text is used as query, only a few keywords are usually provided. In
contrast, when an image is used as query, ``all the information it contains'' is
provided to the system. It is generally said that {``a picture is worth
  a thousand words''} but in the context of information retrieval, which word(s)
is meant when an image is used as a query ? Content based image retrieval (CBIR)
systems attempt to find similar images of an image query from a visual standpoint but in most cases the user is rather
interested in some underlying semantic meanings of the image query.

\begin{figure}
\begin{center}
\includegraphics[width=0.75\textwidth]{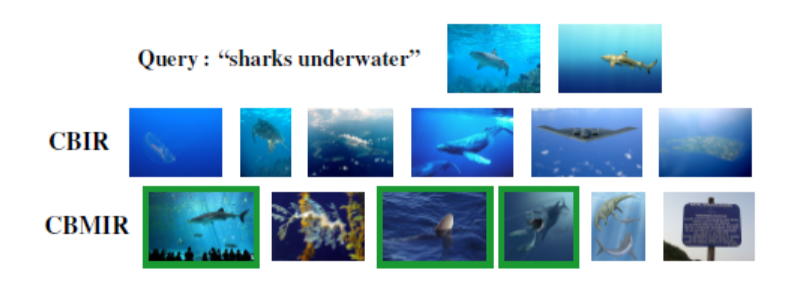}
\end{center}
\caption{Top retrieved images with CBIR and CBMIR for the topic 22  
at ImageCLEF Wikipedia Challenge 2010.}
\label{fig:shark}
\end{figure}

To illustrate this ambiguity, let us consider the example given in Figure \ref{fig:shark}. With regard to this topic, we provide the results obtained with a CBIR system which only uses an image query and also the ones provided by a CBMIR system which uses both visual and textual information. If we only use the image query, we can see that the CBIR system retrieves visually similar images but these latter items are in fact irrelevant to the information need of the user. Indeed, the text query associated with this topic is ``shark underwater''. The images retrieved by the CBIR system show a blue background with, most of them, objects with a fish-like shape. However, none of these images contains a shark. There is a semantic mismatch between the user's information need and the images retrieved by the CBIR system. On the contrary, the images retrieved by the CBMIR system contain sharks for most of them even if their visual similarity values with the image query are lower than the ones given by the CBIR system. As a result, the list retrieved by the CBMIR system is more relevant because the text query gives the semantic meaning of the images the user is interested in unlike the CBIR system. 

 In this paper, we apply the semantic filtering method described in \cite{Clinchant_etal_icmr11}. 
 This filtering operator aims at semantically {correcting} visual similarities between two multimedia items by using their textual similarities and this approach already showed to lead to better results in CBMIR. The semantic operator defined in \cite{Clinchant_etal_icmr11} amounts to filtering visual scores as follows~:
\begin{eqnarray}\label{eq.filtering_sv}
s_{v}^{q_t}(q,d)=
\left\{
\begin{array}{ll}
 s_v(q,d) & \text{if $d$ is in the top $l$ list according to $s_t(q,.)$} \\
0 & \text{otherwise}
\end{array}
\right. 
\end{eqnarray}
The number of selected documents we consider (the nearest neighbors) is equal to $l = min(nnz(s_t(q,.)),m))$, where $nnz(s_t(q,.))$ corresponds to the number of non  zero textual scores given the query $q$, and $m$ is a maximum number of documents which are considered to be pseudo-relevant ($m=1,000$ in our experiments). 

If image reranking \cite{UNED_2010,Popescu10} implicitly uses a similar approach, filtering is essentially thought of as
a pragmatic way to {combine} text and image scores, as the text is only used to
select the documents to be ranked and the ranking is done by the visual scores.
Here, we adopt the view of \cite{Clinchant_etal_icmr11} where the 
filtering step similarly used to select the documents to be considered, but
after the visual scores are recombined with the textual scores for a 
final ranking.  This fundamental difference lead \cite{Clinchant_etal_icmr11} 
to the following method~: after having semantically filtered the visual
similarities $s_v(q,.)$ in order to obtain $s^{q_t}_v(q,.)$, a late fusion 
approach scheme based upon
the weighted mean between $s^{q_t}_{v}(q,.)$ and $s_{t}(q,.)$ was used to rank
and  it was showed that such an approach outperforms other late
fusion methods and also the basic visual reranking method\footnote{Note that the
  method proposed in \cite{Clinchant_etal_icmr11} amounts to linearly combining
  the visual reranking scores and the textual scores, which makes this method
  different from the visual reranking techniques presented in paragraph
  \ref{subsec.vis_rerank}. Indeed in the latter case the top selected 
documents are ranked based on their visual similarity, while the former ranks
the documents based on the fused scores and has been shown to yield 
a much better retrieval performance.}.
Similarly, here we use the filtering to pre-select the documents on which we
apply our relevance models. We argue that there is an inherent ambiguity 
when one  use visual query and filtering is thought of as a way to 
{correct} visual similarities and to specify an information need. We admit
newertheless that this approch has the limitation of ignoring (loosing) 
relevant visual  documents with no textual or irrelevant textual
information. Newertheless, note that our textual filtering step uses text
retrieval techniques that goes beyound simple
keyword matching (see appendix A) and is able to retrieve documents that have
semantic similarity with the query (using\eg lexical entailement and query
expansion).

Hence, in this paper, we extend \cite{Clinchant_etal_icmr11} by using the semantic
filtering strategy in the context of graph based methods. We propose to apply
this filtering scheme to any multimedia scores and similarities before employing
a graph based relevance model. We thus use the top $l$ list given by $s_t(q,.)$
to semantically filter all other similarity matrices and relevance
scores. Indeed, we have previously argued that text based relevance scores are
usually better in retrieving relevant documents in CBMIR. Therefore, we want to
favor the top $l$ list given by $s_t(q,.)$ in any similarities and relevance
scores involved in the fusion process. To this end, we apply the same kind of
semantic filter given in Eq. \ref{eq.filtering_sv} not only to $s_v(q,.)$ but
also to $S_v$, $S_t$ and $s_t(q,.)$ itself. As a consequence, we introduce the
following text query based {\bf semantically filtered visual
  similarities}\footnote{Note that while the retrieval process using such
  similarities is not any more pure visual retrieval, the similarities scores
themselves $S_{v}^{q_t}(d,d')$ are purely visual similarities computed between
the visual signatures of $d$ and $d'$.}~: 
\begin{eqnarray}\label{eq.filtering_Sv}
S_{v}^{q_t}(d,d')=
\left\{
\begin{array}{ll}
 S_v(d,d') & \text{if $d$ and $d'$ are in the top $l$ list according to $s_t(q,.)$} \\
0 & \text{otherwise}
\end{array}
\right. 
\end{eqnarray}

Similarly, we respectively define text query based semantically filtered $s^{q_t}_{t}(q,.)$ and $S^{q_t}_{t}$ as follows~:
\begin{eqnarray}\label{eq.filtering_st}
s_{t}^{q_t}(q,d)=
\left\{
\begin{array}{ll}
 s_t(q,d) & \text{if $d$ is in the top $l$ list according to $s_t(q,.)$} \\
0 & \text{otherwise}
\end{array}
\right. 
\end{eqnarray}

\begin{eqnarray}\label{eq.filtering_St}
S_{t}^{q_t}(d,d')=
\left\{
\begin{array}{ll}
 S_t(d,d') & \text{if $d$ and $d'$ are in the top $l$ list according to $s_t(q,.)$} \\
0 & \text{otherwise}
\end{array}
\right. 
\end{eqnarray}

Therefore, in what follows, the vectors $s^{q_t}_{t}(q,.)$ and
$s^{q_t}_{v}(q,.)$ are sparse and contain only $l\ll n$ non zero
elements. Similarly, each row of the matrices $S^{q_t}_{t}$ and $S^{q_t}_{v}$
only contains $l$ non zero\footnote{The main idea is that given a query the top
  $l$ documents are selected from the collection and the rest of the collection
  is not considered. In terms of  matrix representation  $S^{q_t}_{v}$ 
is a sparsification  of $S_{v}$ where  elements $(i,j)$ of the 
matrix are set to zero except  the ones where both $d_i$ and $d_j$ are amongst
the top $l$ selected ones.  The non-zero elements of $S^{q_t}_{v}$ form an 
$lxl$ sub-matrix of $S^{v}$. In the case of the text the aim is less the 
semantic alignement but the sparsification makes the computational cost
 feasible for large scale datasets.}


Moreover, since we are only interested in the top $l$ list provided by the text relevance scores, we can also remove from $S^{q_t}_{t}$ and $S^{q_t}_{v}$ the rows and columns of items that do not belong to this list. As a consequence, in practice, in order to alleviate the memory and time complexities of graph based techniques, when we compute the relevance and similarity values with respect to a text query $q_t$, we consider $s^{q_t}_{t}(q,.)$ and $s^{q_t}_{v}(q,.)$ as vectors of size $1\times l$ and $S_t$ and $S_v$ as matrices of size $l\times l$. 

Consequently, the semantic filtering approach not only allows one to better bridge the semantic gap but it also dramatically improves the memory complexity since we only need to store matrices of size $O(l^2)$ instead of $O(n^2)$. Furthermore, as we discussed in the previous section, graph based methods rely on diffusion processes which, from an algebraic viewpoint, are materialized by matrix multiplication operations. Since this calculation has a cubic computation complexity with respect to the size of the matrix, the semantically filtered technique enables reducing the time complexity of the graph based methods as well, notably from $O(n^3)$ to $O(l^3)$. Overall, this method makes the graph based techniques scalable for very large multimedia repositories. 
From a more theoretical standpoint, we introduce in the sequel our multimedia relevance model which makes use of the text query based semantic filtering as a core principle and which relies on an unifying framework for graph based techniques that encompasses the methods we have detailed in section \ref{graph_based}.

\section{A unifying framework using semantic filtering and graph based methods}\label{comp_cm_rw}

We now introduce our multimedia retrieval model. Firstly, our framework uses the text query based semantic filtering as a first level of information fusion. In other words, given a text query $q_t$, we start by selecting a subset of semantically relevant items by restraining the search space to the top $l$ elements provided by the textual scores $s_t(q,.)$. In practice, we apply the semantic filters given by Eqs. \ref{eq.filtering_sv}, \ref{eq.filtering_Sv}, \ref{eq.filtering_st} and \ref{eq.filtering_St}. We then propose to apply the graph based methods described in section \ref{graph_based} to the text query based semantically filtered scores and similarities. This leads us to define the following cross-media similarities and derived relevance scores~:
\begin{eqnarray}\label{eq.ctv_qt}
cm^{q_t}_{tv}(q,.)=\mathbf{K}(s^{q_t}_t(q,.),k)\cdot S^{q_t}_v
\end{eqnarray}
\begin{eqnarray}\label{eq.cvt_qt}
cm^{q_t}_{vt}(q,.)=\mathbf{K}(s^{q_t}_v(q,.),k)\cdot S^{q_t}_t
\end{eqnarray}
As argued previously in section \ref{graph_based} we can further fuse the cross-media similarities with semantically filtered textual and visual scores~:
\begin{eqnarray}\label{cm_formula_qt}
rsv^{q_t}_{cm}(q,.)= \alpha_t s^{q_t}_t(q,.)+\alpha_v s^{q_t}_v(q,.)+\alpha_{tv} cm^{q_t}_{tv}(q,.)+ \alpha_{vt} cm^{q_t}_{vt}(q,.)
\end{eqnarray}

In the same manner, we can define retrieval models based on random walks over the stochastic matrix obtained from the semantically filtered multimedia similarities. For the method that takes into account a text based prior given by $s^{q_t}_t(q,.)$, we have~:
\begin{eqnarray}\label{eq.it_rw_tv_qt}
 x^{q_t}_{(i)}=(1-\gamma) x^{q_t}_{(i-1)}\cdot P^{q_t} + \gamma s^{q_t}_t(q,.)
\end{eqnarray}
where $P^{q_t}= D^{q_t} \cdot C^{q_t}$, $C^{q_t}=(1-\beta) S^{q_t}_v+ \beta S^{q_t}_t$ and $D^{q_t}$ is the diagonal matrix whose entries are such that $D^{q_t}(i,i)=1/\sum_{j=1}^nC^{q_t}(i,j)$ and $D^{q_t}(i,j)=0$ for all $i\neq j$. 
The stationary distribution of the previous equation is such
that\footnote{Denoted in \cite{DBLP:conf/mm/HsuKC07} by PRTP}~:
\begin{eqnarray}\label{eq.rw_tv_qt}
 x^{q_t}_{\infty}=(1-\gamma) x^{q_t}_{\infty}\cdot P^{q_t} + \gamma s^{q_t}_t(q,.)
\end{eqnarray}
Then, for the random walk based scores with a prior depending on $s^{q_t}_v(q,.)$, its stationary distribution is given by~:
\begin{eqnarray}\label{eq.rw_vt_qt}
 y^{q_t}_{\infty}=(1-\gamma) y^{q_t}_{\infty}\cdot P^{q_t} + \gamma s^{q_t}_v(q,.)
\end{eqnarray}
Following Eq. \ref{rw_formula}, we can further linearly fused the random walk based scores with $s^{q_t}_t(q,.)$ and $s^{q_t}_v(q,.)$ which yields to~:
\begin{eqnarray}\label{rw_formula_qt}
rsv^{q_t}_{rw}(q,.)=\alpha_t s^{q_t}_t(q,.)+\alpha_v s^{q_t}_v(q,.)+\alpha_{tv} rw^{q_t}_{tv}(q,.)+\alpha_{vt} rw^{q_t}_{vt}(q,.)
\end{eqnarray}

\begin{figure}[t]
\begin{center}
\includegraphics[width=11cm]{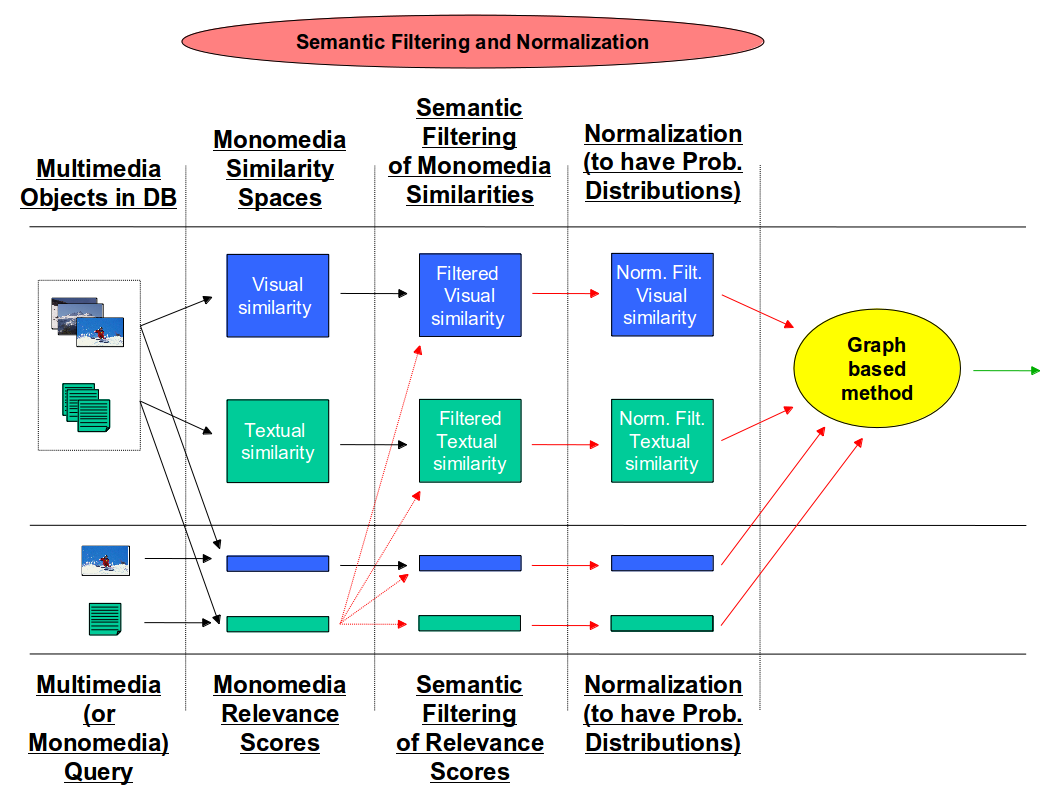}
\caption{Pre-processing, text query based semantic filtering and normalization.\label{fig:preprocessing}}
\end{center}
\end{figure}

In Figure \ref{fig:preprocessing}, we depict the first feature of our multimedia retrieval model which applies the text query based semantic filtering to the query and to the multimedia items of the database. 

The second feature of our multimedia retrieval model aims at defining a unifying framework for graph based methods that encompasses the diffusion processes strategies underlying both the cross-media similarities and the random walk based scores. Note that in order to embed these two approaches in the same model, we assume that all similarities have been normalized so that we manipulate probability distributions. Henceforth, we assume that $s^{q_t}_t(q,.)$, $s^{q_t}_v(q,.)$, and rows of $S^{q_t}_t$ and the ones of $S^{q_t}_v$ have non negative values and that they all sum to one\footnote{This amounts to dividing the row vectors by their $L1$ norms.}. This constraint is due to the random walk method but the cross-media approach does not initially require such a normalization and other possibilities exist. We will come back to this point later on in section \ref{subsec:advantages_cm}. The normalization step occurs just after the text query based semantic filtering and just before applying graph based methods as depicted in Figure \ref{fig:preprocessing}.

To establish our unifying graph based model, let us start by studying the random walk approach a little bit deeper and let us consider the following formula :
\begin{eqnarray}\label{eq.rw_tv_bis_qt}
 x^{q_t}_{\infty}=(1-\gamma) x^{q_t}_{\infty}\cdot P^{q_t} + \gamma x^{q_t}_{\infty}\cdot e \cdot  s^{q_t}_t(q,.)
\end{eqnarray}
where $e$ is the $l\times 1$ vector full of 1.

In the previous equation, the sub-part $x^{q_t}_{\infty}\cdot e$ reduces to $1$ since $x^{q_t}_{\infty}$ is a probability distribution. Therefore Eq. \ref{eq.rw_tv_bis_qt} and Eq. \ref{eq.rw_tv_qt} are strictly equivalent. But, in Eq. \ref{eq.rw_tv_bis_qt}, we can factorize the term $x^{q_t}_{\infty}$ to obtain :
\begin{eqnarray}\label{eq.rw_tv_ter_qt}
 x^{q_t}_{\infty}=x^{q_t}_{\infty}\cdot\left[ (1-\gamma)  P^{q_t} + \gamma e \cdot  s^{q_t}_t(q,.)\right]
\end{eqnarray}
Let us introduce the following matrix of size $l \times l$:
\begin{eqnarray}\label{eq.qtv}
 Q^{q_t}_{tv}=(1-\gamma) P^{q_t} + \gamma e \cdot s^{q_t}_t(q,.)
\end{eqnarray}

Using this matrix, Eq. \ref{eq.rw_tv_ter_qt} can be re-written as $x^{q_t}_{\infty}=x^{q_t}_{\infty}\cdot Q^{q_t}_{tv}$. The solution of this equation is the same as the solution of $(x^{q_t}_{\infty})^\top=(Q^{q_t}_{tv})^\top\cdot (x^{q_t}_{\infty})^\top$ where the right superscript $\top$ states for the transpose operation on vectors and matrices. From the latter relation we see that the stationary probability distribution of the random walk is related to an eigen-decomposition problem \cite{Langville:2005:SEM:1055334.1055396}. Indeed, $x^{q_t}_{\infty}$ is clearly the eigenvector of $(Q^{q_t}_{tv})^\top$ associated to the eigenvalue $1$. Since $Q^{q_t}_{tv}$ is a stochastic matrix, $1$ is the highest eigenvalue. As a result, $x^{q_t}_{\infty}$ is the leading eigenvector of $(Q^{q_t}_{tv})^\top$. One efficient way to compute the leading eigenvector of a square matrix is the power method \cite{Langville:2005:SEM:1055334.1055396}. Thus, in practice, we iterate the following equation until convergence in order to determine $rw^{q_t}_{tv}(q,.)$ :
\begin{equation}\label{eq.rw_tv_pm_qt}
 (x^{q_t}_{(i)})^\top=(Q^{q_t}_{tv})^\top\cdot (x^{q_t}_{(i-1)})^\top
\end{equation}
Since $x^{q_t}_{(0)}$ is a probability distribution then so are the vectors $x^{q_t}_{(i)},i>0$ and $x^{q_t}_{\infty}$ represents the stationary distribution which is proportional to the leading eigenvector of $Q^{q_t}_{tv}$.

Let us now consider the following general formula :
\begin{eqnarray}\label{eq.tv_gen_qt}
 x^{q_t}_{(i)} & \propto & 
\mathbf{K}(x^{q_t}_{(i-1)},k) \cdot \left[ (1-\gamma)
   D^{q_t} \cdot (\beta S^{q_t}_t+ (1-\beta) S^{q_t}_v) + \gamma e \cdot  s^{q_t}_t(q,.)
   \right] \\
x^{q_t}_{(0)} & = & s^{q_t}_t(q,.) \nonumber 
 \end{eqnarray}
$x^{q_t}_{(.)}$ can be interpreted as a generalized diffusion process with a 
text based prior. 
From the previous development (Eqs. \ref{eq.qtv} and \ref{eq.rw_tv_pm_qt}), 
we can see that $rw^{q_t}_{tv}(q,.)$ can be
derived from Eq. \ref{eq.tv_gen_qt}  given by the limit vector
$x^{q_t}_{\infty}$ when we use $k=l$ as $\mathbf{K}(x^{q_t}_{(i-1)},l)$ is 
equivalent to $x^{q_t}_{(i-1)}$ (no operator $\mathbf{K}$ is applied). 
In this case, the random-walk is guaranteed to converge and the limit 
value does not depend on the initialization. 

We can see that the generic case Eq. \ref{eq.tv_gen_qt}  combines the idea of considering only a few nearest 
neighbors in the diffusion process as in the case of the cross-media, 
while doing several iterations (until stability) as in the case of the 
random walk. To avoid confusion, we will refer to this approach as the generalized 
diffusion model and denote it by $gd^{q_t}_{tv}$.

When $k<l$,  we do not have a theoretical guarantee
of the convergence of the diffusion process. However, we have experimentally observed  
that after several iterations the scores became stable. It seems that
the set of top $k$ documents remains unchanged throughout the different iterations. In this case, Eq. \ref{eq.tv_gen_qt} becomes quasi-equivalent to a  power iteration as formalized by Eq. \ref{eq.rw_tv_pm_qt} but with the corresponding reduced graph 
of size $k\times k$. Indeed, we can see that the zeros in 
$\mathbf{K}(x^{q_t}_{(i-1)},k)$ will eliminate from $Q^{q_t}_{tv}$, the 
rows corresponding to the documents not selected by the operator $\mathbf{K}$. 
Concerning the columns corresponding to these documents, while 
they contribute to create  $x^{q_t}_{(i)}$, 
the scores corresponding to them will be ignored  in the next step 
when we will apply the operator 
$\mathbf{K}$ on $x^{q_t}_{(i)}$ (note that we assumed that the set of top 
$k$ documents do not change any more in the iterations).

Note that the generalization of the random walk process with the $\mathbf{K}$ operator 
is new in the literature and we are not aware of any similar work. It is
indeed different from the  Link Reduction by $k$ nearest neighbors  
(PRTP-KNN) proposed in  
\cite{DBLP:conf/mm/HsuKC07}, where the  $k$ nearest neighbors are considered 
for each node in the graph. This can be seen as a sparsification of 
the matrix $S^{q_t}$ where in each row and column
only the $k$ highest values are kept non-zero. We did not considered  
and tested such sparsification in our experiments, as the 
best PRTP-KNN yields the same results as PRTP.

On the other hand, let us now consider $\gamma=0$, which cancels the 
prior given by the text based scores; $\beta=0$, which cancels the text 
based similarity matrix in the convex combination in Eq. \ref{eq.it_rw_tv_qt};
and let us iterate Eq. \ref{eq.tv_gen_qt} only once ($i=1$).
In this particular setting $x^{q_t}_{(1)}$ actually corresponds to 
the cross-media similarity $cm^{q_t}_{tv}(q,.)$ given in Eq. \ref{eq.ctv_qt}.


From these previous observations, we have shown that Eq. \ref{eq.tv_gen_qt} is 
a general graph based approach which generalizes both $rw^{q_t}_{tv}(q,.)$ and
$cm^{q_t}_{tv}(q,.)$ methods.

Similarly, we propose the following formula that allows us to generalize the symmetric relations $rw^{q_t}_{vt}(q,.)$ and $cm^{q_t}_{vt}(q,.)$ :
\begin{eqnarray}\label{eq.vt_gen_qt}
y^{q_t}_{(i)} & \propto & 
\mathbf{K}(y^{q_t}_{(i-1)},k) \cdot \left[ (1-\gamma)
   D^{q_t} \cdot (\beta S^{q_t}_v+ (1-\beta) S^{q_t}_t) 
+ \gamma e \cdot  s^{q_t}_v(q,.)
   \right] \\
y^{q_t}_{(0)} & = & s^{q_t}_v(q,.) \nonumber 
\end{eqnarray}
In that case, $y^{q_t}_{(.)}$ is a generalized diffusion process with a
semantically filtered image based prior. In the right member of
Eq. \ref{eq.vt_gen_qt}, what formally changes as compared to
Eq. \ref{eq.tv_gen_qt}, is the substitution of $t$ by $v$ and \textit{vice
  versa}. However, as stated in the introduction, this formula allows one to
consider the symmetric search scenario that has been less investigated in the
context of the random walk approach. In such a case, we suppose not only a text
query but also an image query and we can thus consider using the random walk
technique for multimedia fusion using a semantically filtered visual based
prior. Indeed in Eq. \ref{eq.vt_gen_qt}, we obtain a random walk based 
technique  biased towards $s^{q_t}_v(q,.)$ that will converge to 
$rw^{q_t}_{vt}(q,.)=y^{q_t}_{\infty}$ given by Eq.~\ref{eq.rw_vt_qt}.


As far as the cross-media based approach is concerned, we obtain the already
defined $cm^{q_t}_{vt}(q,.)$ in Eq. \ref{eq.cvt_qt} 
from the Eq. \ref{eq.vt_gen_qt} by using $\gamma=0$,
which  cancels the prior given by the image based scores; $\beta=1$, which
cancels  the image based similarity matrix in the convex combination and 
we iterate Eq. \ref{eq.vt_gen_qt} only once ($cm^{q_t}_{vt}(q,.)=y^{q_t}_{(1)}$).



This unifying framework encompasses the cross-media similarities and the random
walk based method for CBMIR. Eq. \ref{eq.tv_gen_qt} and Eq. \ref{eq.vt_gen_qt} allow us to have a better understanding of the main differences between these two techniques from a conceptual point of view. However, our proposal suggests more than a simple comparison of those two approaches, it invites to a deeper analysis of what are the key points when using graph based techniques in CBMIR. 

\begin{figure}[t]
\begin{center}
\includegraphics[width=11cm]{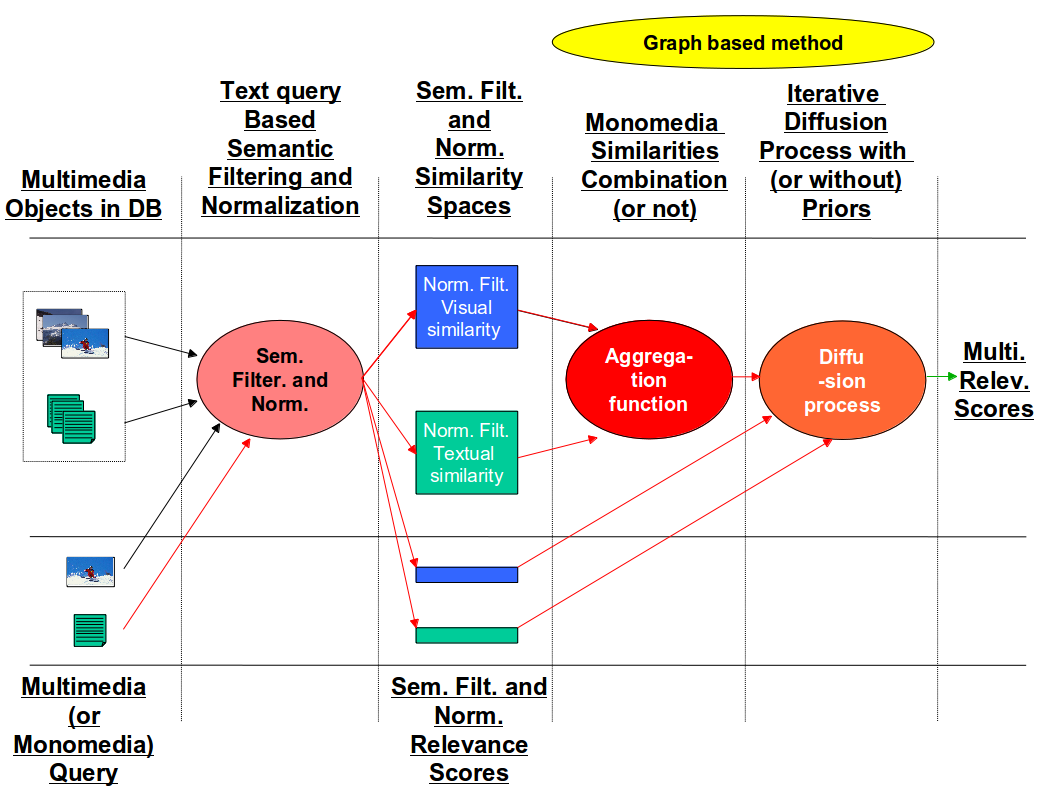}
\caption{Unified view of graph based methods and our multimedia retrieval model.\label{fig:graph}}
\end{center}
\end{figure}

We depict in Figure \ref{fig:graph} the unified formulation of graph based approaches that we have introduced previously accompanied with the preliminary semantic filtering and normalization steps. Overall, this schema represents the multimedia retrieval model we propose in this paper.

In the following sections, we experiment with the proposed multimedia retrieval
model in the case of content based image/text multimedia retrieval. We will
particularly focus on the comparison of three orientations of our framework~:
\begin{itemize}[label=\textbullet]
\item the one that leads to cross-media similarities :
$cm^{q_t}_{tv}=x^{q_t}_{(1)}$ and $cm^{q_t}_{vt}=y^{q_t}_{(1)}$;
\item the one that reduces to random walk based scores :
$rw^{q_t}_{tv}=y^{q_t}_{(\infty)}$, $rw^{q_t}_{vt}=y^{q_t}_{(\infty)}$ 
and $k=l$, meaning that we do not use the operator $\mathbf{K}$;
\item and the generalized diffusion model : $gd^{q_t}_{tv}=y^{q_t}_{(\infty)}$,
$gd^{q_t}_{vt}=y^{q_t}_{(\infty)}$ and $k\ll l$.
\end{itemize}

\section{Experimental settings}\label{subsec.exp_settings}

Firstly, we describe the real-world datasets we applied the different tested techniques to. Then, 
we introduce the image and text representations and similarities 
we used in our experiments. 

\subsection{Datasets}

We conducted our experiments on real-world collections which are constituted of image/text items. The first two datasets were used in the ImageCLEF Photo or Wikipedia retrieval tasks\footnote{\url{http://www.imageclef.org/datasets}} while the last one was constituted in order to assess web image search techniques\footnote{\url{http://lear.inrialpes.fr/~krapac/webqueries/webqueries.html}}. We give below the description of these repositories and the tasks they were meant to address, according to the respective websites that present them.
\begin{itemize}[label=-]
\item The IAPR dataset was used in the context of ImageCLEF 2008 \cite{grubinger:lrec06}. ``The image collection of the IAPR TC-12 Benchmark consists of 20,000 still natural images taken from locations around the world and comprising an assorted cross-section of still natural images. This includes pictures of different sports and actions, photographs of people, animals, cities, landscapes and many other aspects of contemporary life. Each image is associated with a text caption in up to three different languages (English, German and Spanish) . These annotations are stored in a database which is managed by a benchmark administration system that allows the specification of parameters according to which different subsets of the image collection can be generated.''
\item The  Wikipedia collections WIKI10 and WIKI11 were used in ImageCLEF 2010 and 2011 \cite{17013}. ``The Wikipedia image retrieval task is an ad-hoc image retrieval task. The overall goal of the task is to investigate how well multi-modal image retrieval approaches that combine textual and visual evidence in order to satisfy a user’s multimedia information need could deal with larger scale image collections that contain highly heterogeneous items both in terms of their textual descriptions and their visual content. The aim is to simulate image retrieval in a realistic setting, such as the Web environment, where available images cover highly diverse subjects and have highly varied visual properties, while their accompanying textual metadata (if any) are user-generated and correspond to noisy and unstructured textual descriptions of varying quality and length.''\footnote{This is the description of the dataset as provided at \url{http://www.imageclef.org/wikidata}}. Both collections actually contain the same set of 237,434 images. The difference between WIKI10 and WIKI11 is the set of topics used in order to take into account several kinds of multimedia information needs. WIKI10 consists in 70 topics while WIKI11 contains 50 topics. ``The ground truth for these topics was created by assuming binary relevance (relevant \vs non relevant) and by assessing only the images in the pools created by the retrieved images contained in the runs submitted by the participants each year.''
\item The Web Queries (WEBQ) repository was used as a benchmark in order to assess the research work described in \cite{krapac:inria-00548636}. ``The Web Queries dataset contains 71,478 images and meta-data retrieved by 353 web queries. For each retrieved image the relevance label is available. The relevance labels are obtained by manual labeling. French query words were used to retrieve the images, but we provide also the English translation.'' Unlike the previous tasks, WEBQ contains only text topics. Thus, it is a case of asymmetric search scenario.
\end{itemize}

Though we use three different collections, our experiments concern four tasks~:
IAPR, WIKI10, WIKI11 and WEBQ. The tasks are all content based image/text
multimedia data retrieval ones. On each topic given in each task, we tested
different particular cases of the graph based approach introduced in section
\ref{comp_cm_rw}. A topic consists in an image/text query (except for the WEBQ
as explained beforehand) and we were also provided with the binary ground truth
(relevant \vs non relevant). We used the Mean Average Precision (MAP) in order
to compare the obtained rankings and the ground truth in the goal of evaluating
the different multimodal fusion techniques. We also computed 
 if the results were statistically  different using paired t-test at 
the 95\% confidence level.

\subsection{Monomodal Representation and Similarities}
Standard preprocessing techniques were first applied to the textual
part of the documents. After stop-word removal, words were lemmatized 
and the collection of documents indexed with
Lemur\footnote{{http://www.lemurproject.org/}}.
We used a standard Dirichlet language model on IAPR and the Lexical Entailment (LE) information retrieval model \cite{lexical_entailment_ecir06} on the Wikipedia datasets. These models were chosen to remain consistent with 
our previously published and state-of-the-art results \cite{PCGPR10,ap_al_wn08,ap_al_pr09,DBLP:conf/clef/CsurkaCP11,DBLP:conf/clef/ClinchantCAJPSM10}. In fact, the LE model clearly outperforms standard IR models
and give a relative improvement of 15\% MAP\footnote{roughly a raw 4\% in MAP}.
Note that the LE retrieval model is briefly introduced in appendix section \ref{sec:text_models} and was recently rediscovered in \cite{DBLP:conf/sigir/KarimzadehganZ10}.

As for image representations, we used the Fisher
Vector (FV), proposed in \cite{Perronnin07}, an extension of the  popular
 Bag-of-Visual word (BOV) image representation \cite{Sivic03,Csurka04},  
where an image is described by a histogram of quantized local features.
In a nutshell, the Fisher vector consists in modeling the distribution of patches in any image with a Gaussian mixture model (GMM)
 and then in describing an image by its deviation from this average probability distribution.
 In a recent evaluation \cite{Chatfield11}, it has been shown experimentally that the Fisher vector was the state-of-the-art 
 representation for image classification. The Fisher Vector approach is described in appendix section \ref{subsubsec.img_sim}.

 For the purpose of this paper, the choices of a particular textual and visual similarity are not of first importance. Our framework only requires as input a text ranking expert and a visual ranking expert.
So, any textual/visual approaches could be employed and this is why we have moved the descriptions of our experts in the appendix. 
Our focus here is on {the combination} of visual and textual modalities. 
In fact, we did some preliminary  experiments varying the textual and/or the visual features
but the behavior concerning the combination and the conclusions we could draw were the same as for the monomodal experts used in the paper. 
Therefore, they do not bring new insights in our experiments and this is why we did not include these results in this paper.

\section{Experimental results}\label{experiments}

This section contains an extended empirical analysis of the differences between the two graph based methods we are interested in. But in a more general perspective, the experiments we conducted aim at studying the different settings one could apply using the generalization we propose in Eq. \ref{eq.tv_gen_qt} and Eq. \ref{eq.vt_gen_qt}. For convenience, we remind these two principal graph based formulas below~:
\begin{eqnarray*}
x^{q_t}_{(i)} \propto \mathbf{K}(x^{q_t}_{(i-1)},k) \cdot \left[(1-\gamma)
   D^{q_t} (\beta S^{q_t}_t+(1-\beta) S^{q_t}_v) + \gamma e \cdot  s^{q_t}_t
   \right];\,  &  x^{q_t}_{(0)}  =  s^{q_t}_t \\
 y^{q_t}_{(i)} \propto
\mathbf{K}(y^{q_t}_{(i-1)},k)  \cdot \left[(1-\gamma)
    D^{q_t} (\beta S^{q_t}_v+(1-\beta) S^{q_t}_t) + \gamma e \cdot  s^{q_t}_v
   \right];\, & y^{q_t}_{(0)}  =  s^{q_t}_v
\end{eqnarray*}

Our goal is to establish some guidelines on the combination of visual and textual information in CBMIR using graph based methods. To this end, we study several settings of the previously recalled equations and we particularly pay attention to the ones that allow a meaningful comparison between cross-media similarities and random walk based scores.

Accordingly, using Eq. \ref{eq.tv_gen_qt} and Eq. \ref{eq.vt_gen_qt}, we first examine the impact of several parameters on the cross-media and random walk method~:
\begin{itemize}[label=-]
\item What is a good initialization for the graph based methods ?
\item Is it beneficial to iterate the power method until convergence ?
\item What is the impact of the thresholding operator $\mathbf{K}$ ?
\item In which conditions is it beneficial to integrate a text based 
or an image  based  prior in the power method ?
\end{itemize}

Secondly, we investigate on the late combination of the text query based semantically filtered multimedia scores with the graph based scores given by $cm^{q_t}_{tv}$ and $cm^{q_t}_{vt}$ on the one hand, and the random walk based measures $rw^{q_t}_{tv}$ and $rw^{q_t}_{tv}$ on the other hand. In that perspective, we address the following questions~:
\begin{itemize}[label=-]
\item Can we expect benefits from a multimedia query as compared to a text only query ?
\item Is it beneficial to linearly combine the initial semantically filtered
  scores with the ones provided by $cm^{q_t}$, $rw^{q_t}$ or $gd^{q_t}$ ?
\item In which conditions is it beneficial to proceed to a late fusion of similarity matrices before the power method ?
\end{itemize}

All along these empirical analysis, we also comment on the comparison between
the asymmetric and symmetric search scenarios. We recall that in the first case,
only a text query is assumed in order to search the multimedia collection while
in the second case, the user can give an image query in addition to the text
query to better express her information need. 

\begin{figure}[t]
\begin{center}
\includegraphics[width=\textwidth]{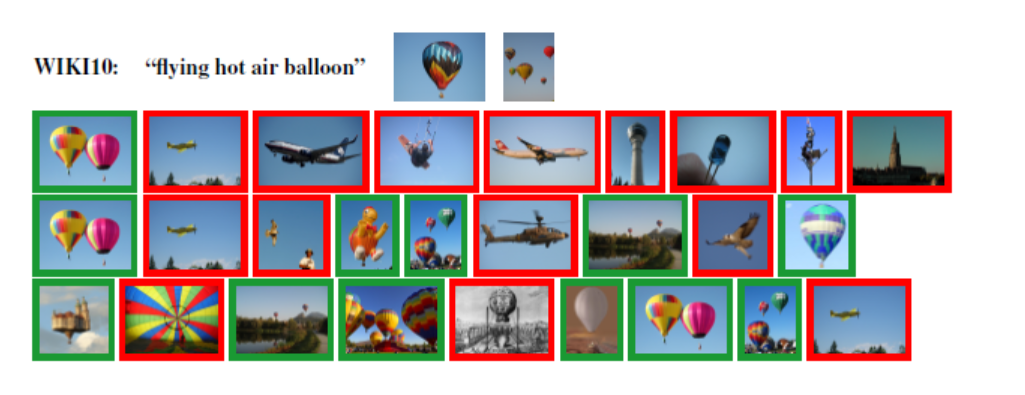}
\end{center}
\caption{Top retrieved images with pure visual similarity (second row), with semantically filtered 
visual similarity  $s^{q_t}_v$  (third row) and with cross-media  $y_{(i)}$ using $k=10$, $\gamma=0$ and $i=1$ (last row), for the topic 9
at ImageCLEF Wikipedia Challenge 2010 (shown in first row). Green means
relevant, red non-relevant. Note that the first two ``non-relevant'' images in
the last row are ``non-flying'' hot air balloons. 
}
\label{fig:baloon}
\end{figure}

Before answering these questions, let us first illustrate three retrieval models for a single query in Figure \ref{fig:baloon}.
This figure shows the beneficial effects of~: a) the text based semantic filtering and b) the advocated cross-media method.

\subsection{Comparison of cross-media similarities and random walk based scores}

First of all, let us recall that $x^{q_t}_{(i)}$ given in Eq. \ref{eq.tv_gen_qt}
and $y^{q_t}_{(i)}$ given in Eq. \ref{eq.vt_gen_qt} are diffusion processes with
priors $s^{q_t}_{t}$ and $s^{q_t}_v$ respectively. As for the initialization of
these iterative equations, $x^{q_t}_{(0)}$ and $y^{q_t}_{(0)}$ could be
 typically set to uniform distributions. However, some preliminary 
results showed that such uniform distributions are suboptimal in the case of
cross-media and does not affect the classical random walk 
(without the operator $\mathbf{K}$). As explained previously to
encompass the cross-media case we consider as initial distributions 
 $s^{q_t}_{t}$ and $s^{q_t}_v$ respectively and 
normalize them  to obtain probability distributions. 


\subsubsection{Impact of the number of iterations $i$}
In our first set of experiments, we vary the number of iterations $i$ in Eq. \ref{eq.tv_gen_qt} and 
Eq. \ref{eq.vt_gen_qt} using the following setting~:
\begin{itemize}
\newcommand{\myitem}{\item[\textbullet]}
\myitem $\gamma=0$ (no prior)
\myitem  $\beta=0$ (no late fusion of similarity matrices)
\end{itemize}
In Table~\ref{tab.rtv_Mq} we show the MAP results we obtained 
when we set $k=l$ (no nearest neighbor operator). In contrast, in 
Table~\ref{tab.rtv_best}, the evaluation measures are shown for 
the best $k$ among $\{1,\ldots,l\}$.  
More precisely, for each task, we first look at the value of $k$ that provided the best MAP measure after the first iteration 
and we then iterated the graph based formulas until convergence with this particular value. 
Best $k$, denoted $k^*$, were in a rather small range (between 10 and 50) for
all tasks (except for WEBQ).

\begin{table}[!t]
\begin{center}
 \begin{tabular}{|c|c|c|c|c|}
\hline
   & IAPR  & WIKI10   & WIKI11 & WEBQ \\
\hline
 \begin{tabular}{c}
\\
\end{tabular} &
 \begin{tabular}{ll}
$s^{q_t}_v$ &  $s^{q_t}_t$ \\
\hline
27.6 & 26.3   
\end{tabular} &
 \begin{tabular}{ll}
$s^{q_t}_v$ &  $s^{q_t}_t$ \\
\hline
 24  & 26.3 
\end{tabular} &
 \begin{tabular}{ll}
$s^{q_t}_v$ &  $s^{q_t}_t$ \\
\hline
18 & 27.8  
\end{tabular} &
 \begin{tabular}{l}
$s^{q_t}_t$ \\
\hline
57 
\end{tabular}\\
\hline
 \begin{tabular}{c}
$i$  \\
\hline
1 \\
2 \\
3 \\
4 \\
5 \\
10 \\
50 \\
$\infty$
\end{tabular} &
 \begin{tabular}{ll}
 $y^{q_t}_{(i)}$ & $x^{q_t}_{(i)}$   \\
\hline
{\bf 28.7} & {\bf 20.8} \\
23.4$^\dagger$ & 17.5$^\dagger$ \\
21.1 & 15.1 \\
19.4 & 13.4 \\
18.7 & 12.3 \\
16.8 & 9.6 \\
15.4 & 8.5 \\
15.4 & 8.4 \\
\end{tabular} &
 \begin{tabular}{ll}
 $y^{q_t}_{(i)}$ & $x^{q_t}_{(i)}$   \\
\hline
{\bf 18.9} & {\bf 15.7} \\
17.2$^\dagger$ & 13.8$^\dagger$ \\
17 & 13.5 \\
17 & 13.5 \\
17 &  13.5 \\
17 &  13.5 \\
17 &  13.5 \\
17 &  13.5 \\
 \end{tabular} &
 \begin{tabular}{ll} 
 $y^{q_t}_{(i)}$ & $x^{q_t}_{(i)}$   \\
\hline
{\bf 12.6} & {\bf 6.9} \\
11.4$^\dagger$ & 5.3$^\dagger$ \\
11.3 & 5.2\\
11.3 & 5.1\\
11.3 & 5.1\\
11.3 & 5.1\\
11.3 & 5.1\\
11.3 & 5.1\\
 \end{tabular} &
\begin{tabular}{l} 
 $x^{q_t}_{(i)}$   \\
\hline
69.3 \\
{\bf 69.5}$^\dagger$  \\
68.7\\
68.4\\
68.4\\
68.4\\
68.4\\
68.4\\
 \end{tabular}\\
\hline
\end{tabular}
\caption{{\bf Varying the number of iterations $i$.} Results obtained 
 with $k=l$ (random walk oriented diffusion process) and $\gamma=\beta=0$. The symbol $\dagger$  indicates a statistical difference between $i=1$ and $i=2$ (which implies a statistical difference between $i=1$ and $i>1$).}
\label{tab.rtv_Mq}
\end{center}
\end{table}

Before focusing on the comparison between cross-media  and 
random walk based results, let us make some preliminary comments~:
\begin{itemize}[label=-]
\item From this first set of experiments, we can observe that the graph based 
scores $x^{q_t}_{(i)}$ and $y^{q_t}_{(i)}$ generally 
 do not outperform the initial scores $s^{q_t}_t$ and $s^{q_t}_v$ 
with respect to MAP values. 
In this first step, we indicate that it is not 
our goal to show  that graph based relevance scores outperform the 
initial semantically 
filtered visual or textual scores.
Our purpose here is rather to compare cross-media based measures against random walk based scores. 

\item Besides, it is interesting to mention that ranking with the 
semantically filtered visual relevance scores  $s^{q_t}_v$ (given 
in Table~\ref{tab.rtv_Mq} or Table~\ref{tab.rtv_best}), lead to 
much better results than ranking with pure image scores
 $s_v$ (without the semantic filtering) since the latter rankings give 
 22.1\%, 6.2\% and 2.7\% for tasks IAPR, WIKI10 and WIKI11 
respectively\footnote{Here we refer
  to the results we obtain when we rank all the documents in the database with
  the visual scores, i.e. no value in $s_v$ is set to zero}. 
As for the WEBQ task, we only have text based queries. 
The superiority of $s^{q_t}_v$ over $s_v$ is particularly true for 
the Wikipedia repository. It is true that the
  direct comparison of  the ranking based on $s_v$ which is pure visual 
with  the ranking based on  $s^{q_t}_v$ which  is multi-modal is nor fair. 
However, it shows that we improving the MAP performance and hence also 
the performance on the top, which is primordial as the top documents 
are used by the operator K in the trans-modal pseudo-relevance step. Indeed
using the textual part of the images in the second row in 
Figure \ref{fig:baloon} to enrich the textual part has better 
chances to improve the results than the textsfrom the images in the  first row.
\item  Furthermore, when the pure visual 
scores are reasonably good (such as for the IAPR task), ranking with
the  semantically filtered visual relevance scores $s^{q_t}_v$ 
(corresponding to the classical visual reranking method) 
outperforms the text based relevance scores $s^{q_t}_t$ too. 
These observations confirm that correcting pure image based 
similarities using text based similarities is beneficial as 
stated in section \ref{sec:sem_filt}. However, as we will see later on,
we can further improve the classical visual reranking results
by  using graph based techniques which provide search 
results that are  complementary. 
\end{itemize}
 
Let us now analyze Tables~\ref{tab.rtv_Mq} and \ref{tab.rtv_best}
 in the goal of comparing the
performances of cross-media similarities and random walk
based scores. Our first core point concerns the number of 
iterations these two approaches assume. Indeed, we recall that when 
$i=1$, the current setting of the parameters of Eq. \ref{eq.tv_gen_qt} 
and Eq. \ref{eq.vt_gen_qt} are respectively similar to $cm^{q_t}_{tv}$ 
and $cm^{q_t}_{vt}$. In contrast, when the number of iterations $i$
grows, the graph based relevance scores converge 
towards $rw^{q_t}_{tv}$ and
$rw^{q_t}_{vt}$ which correspond to the case $i=\infty$. 
The results we obtained enable us to claim that going 
further than a single step in the random walk significantly 
decreases the performances for all tasks, except for WEBQ, 
where a second step was beneficial before the system begun to degrade. 
Hence, we conclude that very short walks give better results than
walking towards convergence and typically, in our case, a one step walk is the 
default setting. These results are to be contrasted with the ones obtained in \cite{Craswell2007}, where the authors found benefits in using long walks\footnote{However, the tasks addressed in \cite{Craswell2007} are different since the graphs they deal with are sparse.}.
Overall, the assumption made by the cross-media method is better than the one underlying the 
random walk technique.  

If we focus on Table~\ref{tab.rtv_best}, we can also compare the cross-media and the generalized diffusion processes. Both methods use the same number of nearest neighbors $k^*$ but the former one makes only one iteration ($i=1$) while the latter one makes iterations until convergence. We can observe that in that case too, very short walks better perform than long walks. 

However, by comparing the last rows of Table~\ref{tab.rtv_best} and Table~\ref{tab.rtv_Mq}, we can conclude that the generalized diffusion process also outperforms the random walk method. Accordingly, it is better to take into account a small set of nearest neighbors in the diffusion process by using the operator $\mathbf{K}$ with $k\ll l$ instead of $k=l$. These experiments suggest that it is better to use short walks and small set of nearest neighbors in our multimedia retrieval model.

\begin{table}[!t]
\begin{center}
 \begin{tabular}{|c|c|c|c|c|}
\hline
   & IAPR  & WIKI10   & WIKI11 & WEBQ \\
\hline
 \begin{tabular}{c}
\\
\end{tabular} &
 \begin{tabular}{ll}
$s^{q_t}_v$ &  $s^{q_t}_t$ \\
\hline
 27.6 & 26.3  
\end{tabular} &
 \begin{tabular}{ll}
$s^{q_t}_v$ &  $s^{q_t}_t$ \\
\hline
24 & 26.3   
\end{tabular} &
 \begin{tabular}{ll}
$s^{q_t}_v$ &  $s^{q_t}_t$ \\
\hline
18 & 27.8  
\end{tabular} &
 \begin{tabular}{l}
$s^{q_t}_t$ \\
\hline
57 
\end{tabular} \\
\hline
 \begin{tabular}{c}
$i$  \\
\hline
1 \\
2 \\
3 \\
4 \\
5 \\
10 \\
50 \\
100 \\
$\infty$\\
\end{tabular} &
 \begin{tabular}{ll}
 $y^{q_t}_{(i)}$ & $x^{q_t}_{(i)}$   \\
\hline
{\bf 35.9} & {\bf 22.4} \\
32.5$^\dagger$ & 20.5$^\dagger$ \\
32.3 & 19.3 \\
32 & 18.3 \\
31.9 & 17.5\\
31.7 & 16.3 \\
31.7 & 15.8 \\ 
31.6 & 15.2 \\
31.6 & 15.2 \\
\end{tabular} &
 \begin{tabular}{ll} 
 $y^{q_t}_{(i)}$ & $x^{q_t}_{(i)}$   \\
\hline
{\bf 25.7} & {\bf 23.9} \\
23.9$^\dagger$ & 21.3$^\dagger$\\
23.4 & 20.3\\
22.8 & 19.9 \\
22.6 & 19.6 \\
22.3 & 19.2 \\
22.1 & 19.1\\
22.1 & 19.1\\
22.1 & 19.1\\
 \end{tabular} &
 \begin{tabular}{ll} 
 $y^{q_t}_{(i)}$ & $x^{q_t}_{(i)}$   \\
\hline
{\bf 21.4} & {\bf 22.5} \\
19.1$^\dagger$ & 18.3$^\dagger$ \\
18 & 16 \\
18 & 15\\
18.3 & 14.5 \\
18.7 & 14.1 \\
18.6 & 14 \\
18.6 & 14 \\
18.6 & 14\\
 \end{tabular} &
\begin{tabular}{l}
$x^{q_t}_{(i)}$   \\
\hline
69.3 \\
{\bf 69.7}$^\dagger$  \\
69.1\\
68.8\\
68.8\\
68.8\\
68.8\\
68.8\\
68.8\\
 \end{tabular} \\
\hline
\end{tabular}
\caption{{\bf Varying the number of iterations $i$.} Results obtained 
 with $k^*$ (best $k$ obtained for the first step $i=1$) (cross-media then generalized oriented diffusion processes) and $\gamma=\beta=0$. 
The symbol $\dagger$ indicates a statistical difference  
between $i=1$ and $i=2$ 
(which implies a statistical difference between $i=1$ and $i>1$).}
\label{tab.rtv_best}
\end{center}
\end{table}

The results shown in Tables~\ref{tab.rtv_Mq} and~\ref{tab.rtv_best}   
also allow us to comment on the different search scenarios. Indeed, in this setting $x^{q_t}_{(i)}$ 
is like having a text query only and the graph based technique propagates
 textual relevance scores to visual similarities. This case is the one that has
 been considered so far in the experiments using random walk based techniques
 and in the research works \cite{DBLP:conf/mm/HsuKC07,DBLP:journals/ieeemm/HsuKC07} in particular. The
 results given by $y^{q_t}_{(i)}$ are, on the contrary, the case where we are
 given in addition to the text query an image query. In that case we use the
 semantically filtered visual relevance scores as a prior and the purpose of
 Eq. \ref{eq.vt_gen_qt} is to propagate the latter measures to text based
 similarities. We observe that $y^{q_t}_{(i)}$ are superior to $x^{q_t}_{(i)}$
 in terms of MAP measures. Hence, these results confirm that, as for
     cross-media similarities, the random walk technique can give better results
     when the user can express her information need by a multimedia query
     instead of a text query solely.

\begin{table}[!t]
\begin{center}
 \begin{tabular}{|c|c|c|c|c|}
\hline
   & IAPR  & WIKI10   & WIKI11 & WEBQ \\
\hline
 \begin{tabular}{c}
\\
\end{tabular} &
 \begin{tabular}{ll}
\hline
$s^{q_t}_v$ &  $s^{q_t}_t$ \\
\hline
 27.6  & 26.3   \\
\end{tabular} &
 \begin{tabular}{ll}
\hline
$s^{q_t}_v$ &  $s^{q_t}_t$ \\
\hline
24 & 26.3   \\
\end{tabular} &
 \begin{tabular}{ll}
\hline
$s^{q_t}_v$ &  $s^{q_t}_t$ \\
\hline
18 & 27.8  \\
\end{tabular} &
 \begin{tabular}{ll}
\hline
 $s^{q_t}_t$ \\
\hline
57  \\
\end{tabular}  \\
\hline
 \begin{tabular}{ccc}
$\gamma$ & $k$   \\
\hline
0 & 10  \\
0.3 & 10 \\
$\gamma^*$  & 10 \\
\hline
0 & 30  \\
0.3 & 30  \\
$\gamma^*$ & 30  \\
\hline
0 & 50  \\
0.3 & 50  \\
$\gamma^*$ & 50 \\
\hline
0 & 100  \\
0.3 & 100   \\
$\gamma^*$ & 100 \\
\hline
0 & $l$   \\
0.3 & $l$   \\
$\gamma^*$ & $l$  \\
\end{tabular} &
 \begin{tabular}{ll} 
 $y^{q_t}_{(i)}$ & $x^{q_t}_{(i)}$   \\
\hline 
35.5 & 19.3$^\dagger$ \\
36.5 & 25 \\
{\bf 36.6} & {26.9}$^\star$ \\
\hline 
34.9 & \mk{22.2}$^\dagger$\\
35.9 & 26.4 \\
35.9  &  {\bf 27.1}$^\star$  \\
\hline 
33.6 & \mk{22.3}$^\dagger$  \\
35.1 & 26.6 \\
35.1 & {26.9}\\
\hline 
\mk{31.3}$^\dagger$  & 21.7$^\dagger$  \\
\mk{33.3} & 26.2 \\
\mk{33.5} & 26.5 \\
\hline 
\mk{28.7}$^\dagger$ & 20.8$^\dagger$\\
\mk{33.4} & \mk{26.6} \\
\mk{33.4} & 26.6 \\
\end{tabular} &
 \begin{tabular}{ll}  
 $y^{q_t}_{(i)}$ & $x^{q_t}_{(i)}$   \\
\hline 
24$^\dagger$ &  23.5$^\dagger$ \\
28.1 & {\bf 29.9}\\
{\bf 28.2} &  {\bf 29.9}\\
\hline 
\mk{25.7}$^\dagger$ & 23.7$^\dagger$ \\
27.9 &  29.8 \\
28 &  29.8 \\
\hline 
25.7$^\dagger$ & 23$^\dagger$ \\
 28 & 29.8 \\
{\bf 28.2} & {\bf  29.9} \\
\hline 
25.2$^\dagger$ & 21.4$^\dagger$ \\
27.4 & 29.5 \\
27.6 & 29.7 \\
\hline 
\mk{18.9}$^\dagger$ & \mk{15.7}$^\dagger$\\
\mk{25.9} & \mk{28.3} \\
\mk{25.9} & \mk{28.3} \\
\end{tabular} &
 \begin{tabular}{ll} 
 $y^{q_t}_{(i)}$ & $x^{q_t}_{(i)}$   \\
\hline 
19.9$^\dagger$ & 22.5$^\dagger$  \\
22.9 & {\bf 31} \\
23.1 & {\bf 31} \\
\hline 
 21.4$^\dagger$ &  \mk{19.9}$^\dagger$ \\
23.2 & 30.6 \\
 23.2 & 30.7 \\
\hline 
21.3$^\dagger$ & \mk{16.4}$^\dagger$ \\
{\bf 23.3} & 30.2\\
{\bf 23.3} & 30.2 \\
\hline 
18.9$^\dagger$ & \mk{12.7}$^\dagger$ \\
22.1 & \mk{29.6}  \\
22.3 & \mk{29.6}  \\
\hline 
\mk{12.6}$^\dagger$ &  \mk{6.9}$^\dagger$ \\
\mk{19.6}  &  \mk{27.8}\\
\mk{19.9} &  \mk{28.2}\\
 \end{tabular} &
\begin{tabular}{ll}
 $x^{q_t}_{(i)}$   \\
\hline 
66.1$^\dagger$ \\
 66.2\\
 67.4$^\star$ \\ 
\hline 
\mk{68.3}$^\dagger$ \\
66 \\
67.9$^\star$\\
\hline 
\mk{68.7}$^\dagger$  \\
66 \\
\mk{68.2}$^\star$\\
\hline 
\mk{69.1}$^\dagger$  \\
66.1 \\
\mk{68.5}$^\star$ \\
\hline
{\bf \mk{69.3}}$^\dagger$  \\
66.1 \\
\mk{68.7}$^\star$ \\
 \end{tabular}\\
\hline
\end{tabular}
\caption{{\bf Varying the number of nearest neighbors $k$ and 
the weight of the prior.} 
Results are shown for $i=1$, $k\in\{10,30,50,100,l\}$ and
  $\gamma\in\{0,0.3, \gamma^*\}$, where $\gamma^*$ was the best $\gamma$
found in the set of $\{0.1,0.2,,\cdots,0.9\}$. 
Adding a prior always leads to significantly better results, 
as also shown by  the symbol $\dagger$ indicating a statistical difference  
between $\gamma=0.3$ (often best or close to best) and $\gamma=0$. 
In contrast, there is rarely a statistical difference between $\gamma=0.3$ 
and $\gamma^*$ indicated by the symbol $\star$. Finally, if there is 
a statistical difference  between $k=10$ and other $k$ values 
the results of $k>10$  are colored in magenta.}
\label{tab.gamma01}
\end{center}
\end{table}

\subsubsection{Impact of the number of nearest neighbor $k$}
In what follows, we
study the results provided by different settings using different values of
$k$. Moreover, we focus on the impact of using a prior or not in
Eqs.~\ref{eq.tv_gen_qt} and \ref{eq.vt_gen_qt}. Indeed, as suggested in
\cite{DBLP:conf/mm/HsuKC07}, it is important to add a prior in order to avoid
the random walk process getting trapped in sub-local optimal 
solutions independent of the query. Hence, we consider the 
following set of parameters~:
\begin{itemize}[label=\textbullet]
\item $k\in\{10,30,50,100,l\}$ (with or without nearest neighbor operator)
\item $\gamma\in\{0,0.3,\gamma^*\}$ (with or without prior, where $\gamma^*$
is the parameter value among $\{0.1,0.2$, $\ldots, 0.9\}$ that gave the 
best performance)
\item $\beta=0$ (no late fusion of similarity matrices)
\end{itemize}
The results using these different settings for $i=1$ are given in
Table~\ref{tab.gamma01} and with $i=\infty$ in 
Table~\ref{tab.gammainfty}. 
Note that the last rows with $k=l$ correspond to 
the classical random walk ($rw^{q_t}$) while the other rows,
with $k \ll l$, lead to the generalized diffusion model ($gd^{q_t}$) that integrates the nearest neighbor operator.


\begin{table}[!t]
\begin{center}
 \begin{tabular}{|c|c|c|c|c|}
\hline
   & IAPR  & WIKI10   & WIKI11 & WEBQ \\
\hline
 \begin{tabular}{c}
\\
\end{tabular} &
 \begin{tabular}{ll}
\hline
$s^{q_t}_v$ &  $s^{q_t}_t$ \\
\hline
 27.6  & 26.3   \\
\end{tabular} &
 \begin{tabular}{ll}
\hline
$s^{q_t}_v$ &  $s^{q_t}_t$ \\
\hline
24 & 26.3   \\
\end{tabular} &
 \begin{tabular}{ll}
\hline
$s^{q_t}_v$ &  $s^{q_t}_t$ \\
\hline
18 & 27.8  \\
\end{tabular} &
 \begin{tabular}{ll}
\hline
 $s^{q_t}_t$ \\
\hline
57  \\
\end{tabular}  \\
\hline
 \begin{tabular}{ccc}
$\gamma$ & $k$   \\
\hline
0 & 10   \\
0.3 & 10 \\
$\gamma^*$  & 10 \\
\hline
0 & 30   \\
0.3 & 30  \\
$\gamma^*$ & 30  \\
\hline
0 & 50   \\
0.3 & 50   \\
$\gamma^*$ & 50 \\
\hline
0 & 100   \\
0.3 & 100   \\
$\gamma^*$ & 100  \\
\hline
0 & $l$ \\
0.3 & $l$   \\
$\gamma^*$ & $l$   \\
\end{tabular} &
 \begin{tabular}{ll} 
 $y^{q_t}_{(i)}$ & $x^{q_t}_{(i)}$   \\
\hline 
31.4$^\dagger$ & 16.6$^\dagger$ \\
35.1 & 24.2 \\
{\bf 35.4} & 26.8$^\star$ \\ 
\hline 
29$^\dagger$ & 14.6$^\dagger$\\
33.4 & 24.1 \\
34.3$^\star$  &  26.9$^\star$  \\ 
\hline 
29.3 & 15.9$^\dagger$ \\
\mk{30.7} & 24.6 \\
33.1 & {\bf 26.7}$^\star$\\ 
\hline 
\mk{24.9}$^\dagger$  & 15.5$^\dagger$  \\
\mk{30.2} & 25.1 \\
32.3 & 26.4 \\ 
\hline 
\mk{15.4}$^\dagger$ & \mk{8.4}$^\dagger$ \\
31.9 & \mk{26.4}\\
32.1 & 26.6 \\ 
\end{tabular} &
 \begin{tabular}{ll}  
 $y^{q_t}_{(i)}$ & $x^{q_t}_{(i)}$   \\
\hline 
22.2$^\dagger$ &  20.7$^\dagger$ \\
28.4 & 29.2 \\
{\bf 28.4} &  29.2\\ 
\hline 
22.1$^\dagger$ & \mk{18.3}$^\dagger$ \\
27.8 &  {\bf 29.4} \\
27.8 &  {\bf 29.4} \\ 
\hline 
19.9$^\dagger$ & \mk{18.1}$^\dagger$ \\
 26.2 & {\bf 29.4} \\
27.3 & {\bf  29.4} \\ 
\hline 
\mk{18.9}$^\dagger$ & \mk{16.6}$^\dagger$ \\
\mk{25.5} & 28.4 \\
\mk{26.3} & 28.4 \\ 
\hline 
\mk{17}$^\dagger$ & \mk{13.5}$^\dagger$ \\
\mk{25.3} & 27.9\\
\mk{25.5} & 27.9 \\
\end{tabular} &
 \begin{tabular}{ll} 
 $y^{q_t}_{(i)}$ & $x^{q_t}_{(i)}$   \\
\hline 
19.1$^\dagger$ & 14$^\dagger$  \\
{\bf 22.8} & {\bf 31.3} \\
{\bf 23.6} & {\bf 31.3} \\ 
\hline 
18.6 &  \mk{10.2}$^\dagger$ \\
22.1 & 30.9 \\
 22.4 & 30.9 \\ 
\hline 
15.1$^\dagger$ & \mk{8.7}$^\dagger$ \\
20.3 & 29.4\\
 21 & 30 \\ 
\hline 
\mk{10.8}$^\dagger$ & \mk{7.2}$^\dagger$ \\
\mk{17.3} & \mk{28.5}  \\
20.5 & \mk{28.9}  \\ 
\hline 
\mk{11.3}$^\dagger$ & \mk{5.1}$^\dagger$ \\
19.1 & \mk{27.6} \\
\mk{19.8} & \mk{28.2} \\
 \end{tabular} &
\begin{tabular}{ll}
 $x^{q_t}_{(i)}$   \\
\hline 
66.4 \\
 66\\
 66.9$^\star$\\
\hline 
69.1 \\
67.3 \\
69.7\\ 
\hline 
69.5$^\dagger$  \\
67.7 \\
{\bf 70.4}$^\star$\\  
\hline 
69.3$^\dagger$  \\
67.8 \\
70.3$^\star$ \\
\hline 
\mk{68.4}$^\dagger$   \\
\mk{66.6} \\
\mk{69.5}$^\star$\\
 \end{tabular}\\
\hline
\end{tabular}
\caption{{\bf Varying the number of nearest neighbors $k$ and 
the weight of the prior.} 
Results are shown for $i=\infty$, $k\in\{10,30,50,100,l\}$ and
  $\gamma\in\{0,0.3, \gamma^*\}$, where $\gamma^*$ was the best $\gamma$
found in the set of $\{0.1,0.2,,\cdots,0.9\}$. 
Adding a prior always leads to significantly better results, 
as also shown by  the symbol $\dagger$ indicating a statistical difference  
between $\gamma=0.3$ (often best or close to best) and $\gamma=0$. 
In contrast, there is rarely a statistical difference between $\gamma=0.3$ 
and $\gamma^*$ indicated by the symbol $\star$ (except WEBQ where we always found
 that the best $\gamma$ was $0.1$). Finally, if there is 
a statistical difference  between $k=10$ and other $k$ values 
the results of $k>10$  are colored in magenta.}
\label{tab.gammainfty}
\end{center}
\end{table}

Analyzing these  results, and excluding the case of WEBQ, 
we observe the followings~:
\begin{itemize}[label=-]
\item Concerning the $k$ value, best or near best results are obtained with 
$k=10$ for any value of $\gamma$. When the best results are achieved with 
$k>10$, the latter parameter is below or equal to $50$ and the gain is neither high nor
statistically different as compared to $k=10$. Accordingly, we conclude that for both the
cross-media and the generalized diffusion model using $k \approx 10$   
could be considered as a default setting of our multimedia relevance model. 
\item When $\gamma=0$, the $cm^{q_t}$ approach with $i=1$ (shown in Table~\ref{tab.gamma01}),
 always yields to much higher performances than the $gd^{q_t}$ method with $i=\infty$ (shown in Table~\ref{tab.gammainfty}). While this is not 
always the case when we have $\gamma>0$, the setting $\gamma^*$, $k=10$ 
and $i=1$ yields to results that are close (and statistically not different) 
to the best obtained values. In other words, and as already suggested beforehand, when using graph based techniques, 
one should favor the cross-media oriented diffusion process along 
with a relatively small number of nearest neighbors as proxies.
\item When we consider the random walk, it generally yields  
to worse results than $cm^{q_t}$ and $gd^{q_t}$ for all three datasets whatever the value of $\gamma$.  
 \end{itemize}

Note, that in the  case of WEBQ, some of  these observations seems to 
be less true,  where increasing both  $k$ and $i$ improves the 
search results in terms of MAP.  Best results are obtained with $k=50$,
$\gamma=0.1$ and  $i=\infty$, which again shows that using the operator 
$\mathbf{K}(.,k)$ (but this time with a higher $k$) is a good idea. Concerning
 the number of iterations,  we obtain a nice improvement over 
$i=1$ and close to  $i=\infty$ with a single extra step\footnote{With 
$\gamma^*=0.1$,  we  obtain  that MAP=$\{69.8,69.9$, $69.8,69.8,69.7\}$ 
respectively for
$k\in\{10,30,50,100,l\}$}. This shows that even if 
a single iteration is not the best option, using only few iterations 
and small $k$ values are sufficient. This confirms the tendencies we have observed so far.

\subsubsection{Impact of  the prior weight $\gamma$}

Next, we can observe the following facts in regard to the integration of a prior in the graph based methods ($\gamma=0$ 
\vs $\gamma>0$.)  The case $\gamma>0$, $k=l$, $i=\infty$, is typical to 
the initial random walk technique suggested in \cite{DBLP:conf/mm/HsuKC07,DBLP:journals/ieeemm/HsuKC07}. In contrast, the case $\gamma=0$, $k\leq l$, $i=1$,  is related to the cross-media technique. However, setting $\gamma>0$ in the latter case adds a prior in the cross-media oriented diffusion process and such a model has not been evaluated so far.

From the three last rows in Table~\ref{tab.gammainfty}, we can deduce that adding a prior 
to the random walk approach ($i=\infty$) generally improves the search results. 
This outcome is aligned with the findings in \cite{DBLP:conf/mm/HsuKC07,DBLP:journals/ieeemm/HsuKC07} 
which showed that {{the prior towards textual relevance scores allows better search results}}. 
In the case of a symmetric search scenario, we can also compute $y^{q_t}_{(i)}$ and in that case 
too, adding a semantically filtered visual prior improves the random walk based diffusion process performances. 
We thus extend the findings provided in  \cite{DBLP:conf/mm/HsuKC07,DBLP:journals/ieeemm/HsuKC07} to the symmetric 
search scenario. 

Yet, $y^{q_t}_{\infty}$ scores do not surpass $x^{q_t}_{\infty}$ ones except for the IAPR task. Accordingly, using an image query with the random walk technique does not necessarily improve the search results. This observation might partly explain why research works using random walks in multimedia fusion have not studied further the symmetric search scenario. Note nevertheless that the random walk ($y^{q_t}_{\infty}$) always outperforms the  image reranking ($s^{q_t}_v$) score.

If we compare  the effect of adding a prior in the cross-media oriented diffusion process ($i=1$), by setting $\gamma>0$, we observe that in all cases introducing priors was  beneficial. 
Concerning the weight $\gamma$,  we found that in most cases $\gamma=0.3$ was
 either optimal or yielded to values close to optimal, except in the case of the
 WEBQ dataset, where  $\gamma=0.1$ leaded to much better results. Therefore, in
 what follows we will set $\gamma$ to 0.3 for IAPR and WIKI datasets and to 0.1
 for WEBQ, and denote these respective settings by $\hat{\gamma}$.

Similarly, adding the prior to the generalized diffusion model $gd^{q_t}$
was always beneficial. Besides, while without the prior ($\gamma=0$) the 
$cm^{q_t}$ models outperform the $gd^{q_t}$ ones, this is less true when we set $\gamma>0$. Indeed, if we add a prior to both techniques, we observe that the 
generalized diffusion models give similar results than the cross-media models and in some cases (WIKI11), we can even note 
some better results.

At this point of our experimental report, we can make the following comments on
the proposed multimedia retrieval model~: 
\begin{enumerate}[label= (\roman*)]
\item The results given in
Tables ~\ref{tab.gamma01} and \ref{tab.gammainfty} support the fact that graph based methods are
    effective techniques to combine visual and textual information in CBMIR as
    long as the parameters of Eq. \ref{eq.tv_gen_qt} and Eq. \ref{eq.vt_gen_qt}
    are set correctly since we can obtain much better results than the
baselines given by $s^{q_t}_t$ and $s^{q_t}_v$ respectively; 
\item  Adding a prior in any of the presented diffusion process improved the
  results as compared to the ones that do not use any prior.
\item The diffusion processes using nearest neighbors as proxies 
($cm^{q_t}$ and $gd^{q_t}$) outperform  the random walk oriented diffusion 
process for both $y^{q_t}_{(i)}$ and $x^{q_t}_{(i)}$. We also claim that $k \approx 10$ and $\gamma=0.3$ can be considered as 
default parameters values for Eq. \ref{eq.tv_gen_qt} and Eq. \ref{eq.vt_gen_qt}.
\end{enumerate}

Finally, Table \ref{tab.gamma01} allows us to further compare the 
results obtained in an asymmetric search scenario $x^{q_t}_{(i)}$ 
against the ones obtained in a symmetric search scenario 
$y^{q_t}_{(i)}$. Interestingly, even if we use the best settings, 
the MAP measures for $y^{q_t}_{(i)}$ do not always outperform the ones 
for $x^{q_t}_{(i)}$ (actually it outperforms only for IAPR). However, 
these observations do not mean that in CBMIR, multimedia queries are 
not necessarily useful and adding an image query to a text query 
does not help the search results significantly. Indeed, retrieved 
lists provided by $y^{q_t}_{(i)}$ and $x^{q_t}_{(i)}$ should not 
be considered as in competition with each other. They should be viewed 
as complementary lists of pseudo-relevant items and in that perspective, 
$s^{q_t}_v$ and $s^{q_t}_t$ too have to be considered as complementary 
pseudo-relevant top lists of multimedia objects. In the section 
\ref{sec:combscores}, 
we will thus investigate the combination of those different scores 
in order to analyze to what extent these top lists are mutually complementary.

\begin{table}[ttt]
\begin{center}
 \begin{tabular}{|c|c|c|c|c|}
\hline
   & IAPR  & WIKI10   & WIKI11 \\
\hline
 \begin{tabular}{c}
\\
\end{tabular} &
 \begin{tabular}{cc}
\hline
$s^{q_t}_v$ &  $s^{q_t}_t$ \\
\hline
 27.6 & 26.3   \\
\end{tabular} &
 \begin{tabular}{cc}
\hline
$s^{q_t}_v$ &  $s^{q_t}_t$ \\
\hline
 24 & 26.3   \\
\end{tabular} &
 \begin{tabular}{cc}
\hline
$s^{q_t}_v$ &  $s^{q_t}_t$ \\
\hline
18 & 27.8  \\
\end{tabular} \\
\hline
 \begin{tabular}{ccc}
$\beta$  & $i$ & $k$  \\
\hline
0 & 1 & 10 \\
0 & $\infty$ & 10 \\
0 & $\infty$ & $l$\\  
\hline
0.5 & 1 & 10 \\
0.5 & $\infty$ & 10 \\
0.5 & $\infty$ & $l$  \\  
\hline
$\beta^{*}$ & 1 & 10 \\
$\beta^{*}$ & $\infty$ & 10\\  
$\beta^{*}$ & $\infty$ & $l$ \\  
\hline
1 & 1 & 10\\
1 & $\infty$ & 10 \\
1 & $\infty$  & $l$\\
\end{tabular} &
 \begin{tabular}{ll} 
 $y^{q_t}_{(i)}$ & $x^{q_t}_{(i)}$   \\
\hline
36.5 & 25 \\ 
35.1 & 24.2 \\ 
31.9$^\star$ & 26.4$^\star$ \\  
\hline  
36.2 & {\bf \mk{27.3}} \\ 
34 & \mk{27.2} \\
30.7$^\star$ & 26 \\  
\hline
{\bf 36.6} & {\bf \mk{27.3}} \\  
34.9 & \mk{27.2} \\ 
31.7$^\star$ & 26.4 \\  
\hline
\mk{27.7}  &   \mk{27.1} \\ 
\mk{26.3} & 25.9 \\
\mk{28.1}$^\star$ &  26 \\  
\end{tabular} &
 \begin{tabular}{ll} 
 $y^{q_t}_{(i)}$ & $x^{q_t}_{(i)}$   \\
\hline
28.1 &  {\bf 29.9} \\ 
28.4 & 29.2 \\
25.3$^\star$ &  27.9$^\star$ \\  
\hline          
27.1 & 29.3 \\ 
27.3 & 29.2 \\
24.6$^\star$ & 28$^\star$ \\  
\hline
28.5$^\star$ & {\bf 29.9}$^\star$ \\  
\bf{29.2} & 29.3 \\ 
25.5 &  28 \\  
\hline
\mk{24.5}$^\star$ &   \mk{26.6} \\ 
\mk{24.4} & \mk{26.3} \\
\mk{22.7} & \mk{26.7} \\  
\end{tabular} &
 \begin{tabular}{ll} 
 $y^{q_t}_{(i)}$ & $x^{q_t}_{(i)}$   \\
\hline
22.9  & 31  \\  
22.8 & {\bf 31.3} \\
19.1$^\star$ & 27.6$^\star$  \\ 
\hline          
\mk{20.5}  & \mk{29.7}  \\  
21 & \mk{29.5} \\
\mk{15.8}$^\star$ & 27.8$^\star$ \\ 
\hline
 23 & 30.9\\  
{\bf \mk{24}} & 31.2 \\
18.8$^\star$ &  27.8$^\star$  \\  
\hline
\mk{16.6} &  \mk{27.2} \\  
\mk{15.2}$^\dagger$ & 27 \\
\mk{11.5}$^\star$  & 27.1 \\ 
 \end{tabular}\\
\hline
\end{tabular}
\caption{{\bf Combining the similarity matrices.} We vary $\beta$ and 
show the obtained results for $x^{q_t}_{(i)}$ and $y^{q_t}_{(i)}$
 with the prior $\gamma=0.3$. 
The symbol $\dagger$ indicates a statistical difference between them. 
When there is a statistical  difference between $\beta=0$ and $\beta>0$ 
(all the other parameters 
remaining the same), the values of the results given by $\beta>0$  
are  colored in magenta.}
\label{tab.rw_beta}
\end{center}
\end{table}

\subsubsection{Combining the similarity matrices}
We now examine if fusing semantically filtered visual and textual
similarity matrices by setting $\beta\neq 0$ in Eq. \ref{eq.tv_gen_qt} and
Eq. \ref{eq.vt_gen_qt} is beneficial or not. 
Combining monomedia similarity matrices was indeed 
suggested in \cite{DBLP:conf/mm/HsuKC07,DBLP:journals/ieeemm/HsuKC07} 
in the random walk case. Consequently, we study the impact of $\beta$ 
in the following experiments. As a result of the 
previous experiments, we selected the following parameter values~:
\begin{itemize}
\newcommand{\myitem}{\item[\textbullet]}
\myitem $i=1, k=10$ (cross-media oriented diffusion process) 
\myitem $i=\infty, k=10$ (generalised diffusion model)
\myitem $i=\infty, k=l$ (random walk oriented diffusion process)
\end{itemize}
For each case, we set  $\gamma=0.3$ and 
we varied  $\beta\in[0,1]$, but we only show
three particular cases~: $\beta=0$, $\beta=1$ and $\beta^*\in[0.1,0.9]$ 
which correspond to the values that gave the best results for each  model.
Table~\ref{tab.rw_beta} presents the different MAP values we obtained. Note
that we do not show the results for the WEBQ task because we did not 
compute\footnote{As we shall see, setting $\beta>0$ does not necessarily bring any improvements of the results for IAPR, WIKI10 and WIKI11. Therefore, since the computation of visual similarities in the case of WEBQ is heavy, we did not compute it.} 
the visual similarity matrix $S^{q_t}_v$ in this case. 
From Table~\ref{tab.rw_beta} we can make the following observations~:
\begin{itemize}[label=-]
\item Adding $S^{q_t}_v$ to $S^{q_t}_t$ ($\beta >0$) was 
clearly beneficial 
only in a few cases, more precisely for  $cm^{q_t}_{tv}$ and $gd^{q_t}_{tv}$
in the case of  IAPR and for $cm^q_{vt}$ and $gd^q_{vt}$ in 
the case of WIKI11. However, 
in the latter cases, an equal  weighting is far from optimal or even worse
and the parameter $\beta$ has to be properly set. 
As a consequence, we can say that, in general, $\beta =0$ gives the best 
or near optimal solutions and adding the similarity matrix of the complementary
 modality in Eqs. \ref{eq.tv_gen_qt} and \ref{eq.vt_gen_qt} does not bring any
 further gain. 
\item As previously,  both the 
cross-media and the  generalized diffusion model results 
improve over the baselines  $s^{q_t}_v$ and $s^{q_t}_t$ scores, and remain 
significantly better than the classical random walk process.
\end{itemize}

Note that the particular case $\beta=1$ corresponds, either to a 
semantically filtered ``monomodal''
pseudo-relevance feedback ($i=1$),  or to a semantically filtered 
``monomodal'' diffusion process  ($i=\infty$). Compared to 
the opposite case, $\beta=0$, which is a transmodal 
semantically filtered pseudo-relevance 
feedback, they perform in general worse. This actually shows again
the interest of exploiting both modalities using graph based 
techniques with a transmedia principle.

\subsection{Combination of different relevance scores and 
similarities in a late  fusion scheme}
\label{sec:combscores}

In the next set of experiments, we follow the research works described in
\cite{Clinchant_wn07,ap_al_wn08,ap_al_wn09,Clinchant_etal_icmr11} where the
authors propose to combine text based scores, semantically filtered visual
scores and graph based scores as for the final relevance scores. In that perspective, we first study 
the following cases~:
\begin{eqnarray}
rsv_{cm}^{q_t,tv}(q,.)&=&\alpha s^{q_t}_t(q,.)+(1-\alpha)x^{q_t}_{(1)}\label{eq:cmtv}\\
rsv_{cm}^{q_t,vt}(q,.)&=&\alpha s^{q_t}_v(q,.)+(1-\alpha)y^{q_t}_{(1)} \label{eq:cmvt}\\
rsv_{gd}^{q_t,tv}(q,.)&=&\alpha s^{q_t}_t(q,.)+(1-\alpha)x^{q_t}_{\infty} \label{eq:best_rwtv}\\
rsv_{gd}^{q_t,vt}(q,.)&=&\alpha s^{q_t}_v(q,.)+(1-\alpha)y^{q_t}_{\infty} \label{eq:best_rwvt}
\end{eqnarray}

\begin{table}[t]
\begin{center}
 \begin{tabular}{|c|c|c|c|c|}
\hline
   & IAPR  & WIKI10   & WIKI11 & WEBQ\\
\hline
 \begin{tabular}{c}
\\
\end{tabular} &
 \begin{tabular}{ll}
$s^{q_t}_v$ &  $s^{q_t}_t$ \\
\hline
 27.6 & 26.3   \\
\end{tabular} &
 \begin{tabular}{ll}
$s^{q_t}_v$ &  $s^{q_t}_t$ \\
\hline
24 & 26.3   \\
\end{tabular} &
 \begin{tabular}{ll}
$s^{q_t}_v$ &  $s^{q_t}_t$ \\
\hline
18.1 & 27.8  \\
\end{tabular} &
 \begin{tabular}{l}
 $s^{q_t}_t$ \\
\hline
57  \\
\end{tabular}
\\
\hline
 \begin{tabular}{ccc}
 $k$  & $\alpha$ & $\gamma$\\
\hline
 10  & 0  & 0\\
 10  & 0.5 & 0\\
 10  & $\alpha^*$ & 0\\
\hline
 10  & 0  &  $\hat{\gamma}$\\
 10  & 0.5 &  $\hat{\gamma}$\\
 10  & $\alpha^*$ & $\hat{\gamma}$\\
\hline
 30  & 0 & 0 \\
 30  & 0.5 & 0 \\
 30  & $\alpha^*$ & 0 \\
\hline
 30  & 0 & $\hat{\gamma}$ \\
 30  & 0.5 & $\hat{\gamma}$ \\
 30  & $\alpha^*$ & $\hat{\gamma}$ \\
\hline
50 & 0  & 0 \\
 50  & 0.5 & 0\\
 50  &  $\alpha^*$ & 0\\
\hline
50 & 0 & $\hat{\gamma}$ \\
 50  & 0.5 & $\hat{\gamma}$ \\
 50  &  $\alpha^*$ & $\hat{\gamma}$ \\
\hline
 $l$  & 0 & 0\\
 $l$  & 0.5 & 0\\
 $l$  & $\alpha^*$ & 0\\
\hline
 $l$  & 0 & $\hat{\gamma}$ \\
 $l$  & 0.5 & $\hat{\gamma}$ \\
 $l$  & $\alpha^*$ & $\hat{\gamma}$ \\
\end{tabular} &
 \begin{tabular}{ll} 
 $rsv_{cm}^{q_t,vt}$ & $rsv_{cm}^{q_t,tv}$   \\
\hline
35.5 & 19.3$^\dagger$ \\
35.6 & {\bf 27} \\
36.5 & {\bf 27}\\
\hline
36.5 & 25$^\dagger$ \\
\mk{34.5} & {\bf 27}\\
{\bf 36.6}$^\star$ & {\bf 27} \\
\hline  
34.9 & 22.2$^\dagger$ \\
35.1 & 26.7 \\
35.8$^\star$ & 26.9 \\
\hline
35.9$^\dagger$ & \mk{26.4} \\
\mk{33.8} & 27 \\
35.9$^\star$ & 27 \\
\hline 
33.6 & 22.3$^\dagger$ \\
34.5 & 26.6  \\
34.9 & 26.9  \\
\hline
35.1$^\dagger$ &  \mk{26.6}$^\dagger$\\
\mk{33.3} & 26.9 \\
35.1$^\star$ & 26.9 \\
\hline 
28.7$^\dagger$ & 20.8$^\dagger$ \\
33.3 & 25.8 \\
33.3 & 26.7 \\
\hline
\mk{33.4}$^\dagger$ &  \mk{26.6}\\
\mk{31.2} & 26.6 \\
33.4$^\star$ & 26.7 \\
\end{tabular} &
 \begin{tabular}{ll} 
 $rsv_{cm}^{q_t,vt}$ & $rsv_{cm}^{q_t,tv}$   \\
\hline
24$^\dagger$ & 23.5$^\dagger$ \\
28 &  30 \\
28.2 &  30 \\
\hline
\mk{28.1}$^\dagger$ & \mk{29.9}$^\dagger$ \\
\mk{26.9} & \mk{28.8} \\
28.2$^\star$ & 29.9$^\star$ \\
\hline 
25.7$^\dagger$ & 23.7$^\dagger$ \\
27.8 &  {\bf 30.1}\\
28.2$^\star$ & {\bf 30.1} \\
\hline
\mk{27.9}$^\dagger$  & \mk{29.8}$^\dagger$ \\
\mk{26.5} & \mk{28.6} \\ 
\mk{27.9}$^\star$ & 29.8$^\star$ \\ 
\hline 
25.7$^\dagger$ & 23$^\dagger$ \\
27.7 &  30 \\
{\bf 28.3} & 30 \\
\hline 
\mk{28}$^\dagger$ & \mk{29.8}$^\dagger$ \\
\mk{26.4} & \mk{28.5} \\
28$^\star$ & 29.8$^\star$ \\
\hline 
18.9$^\dagger$ & 15.7$^\dagger$ \\
25.9 & 28 \\
25.9 & 28.3 \\
\hline 
\mk{25.9} & \mk{28.3} \\
25.3 & 27.7 \\
25.9 & 28.3 \\
\end{tabular} &
 \begin{tabular}{ll} 
 $rsv_{cm}^{q_t,vt}$ & $rsv_{cm}^{q_t,tv}$   \\
\hline
19.9$^\dagger$  & 22.5$^\dagger$  \\
22.7 & 30.9\\
23.1 & 30.9 \\
\hline
\mk{22.9}  & {\bf \mk{31}}$^\dagger$ \\
\mk{21.8} & 30  \\
23.1$^\star$ & {\bf 31}$^\star$ \\
\hline 
21.4 & 19.9$^\dagger$ \\
22.9 &  30.7 \\
23.3 &  30.7\\
\hline
\mk{23.2}$^\dagger$ & \mk{30.6}$^\dagger$ \\
\mk{21.7} & 29.6 \\
23.3$^\star$ & 30.6$^\star$ \\
\hline 
21.3 & 16.4$^\dagger$ \\
23.1 & 30.3 \\
{\bf 23.5}$^\star$ & 30.3 \\
\hline
23.3$^\dagger$ & \mk{30.2}$^\dagger$ \\
\mk{21.8} & 29.3 \\
\mk{23.4}$^\star$ & 30.2$^\star$ \\
\hline 
12.6$^\dagger$ & 6.9$^\dagger$ \\
19.6 & 26.8 \\
19.7$^\star$ & 28.2$^\star$ \\
\hline
19.6 & 27.8 \\
19.6 & 28.2 \\
19.6 & 28.2 \\
 \end{tabular} &
\begin{tabular}{l}
 $rsv_{cm}^{q_t,tv}$   \\
\hline
64.4$^\dagger$\\
66.9\\
66.9\\
\hline
\mk{66.9}$^\dagger$ \\
\mk{66.2}\\
66.9$^\star$\\
\hline 
68.3 \\
67.9\\
68.6$^\star$\\
\hline
 67.9$^\dagger$\\ 
66.5 \\
67.9$^\star$\\
\hline 
68.7$^\dagger$ \\ 
68.3\\
69$^\star$ \\
\hline
\mk{68.2}$^\dagger$  \\
\mk{66.6}\\
\mk{68.2}$^\star$\\
\hline 
69.3\\
68.9 \\
{\bf 69.6}\\
\hline
\mk{68.7} \\ 
\mk{66.9}\\
\mk{68.7}\\
 \end{tabular} \\
\hline
\end{tabular}
\caption{{\bf Effect of adding the initial semantically filtered 
scores $s^{q_t}_v$ and $s^{q_t}_t$ to the cross-media scores.}  
We show the results  of $rsv_{cm}^{q_t,vt}$ and $rsv_{cm}^{q_t,tv}$ 
with $\beta=0$. The symbol $\dagger$ indicates a statistical 
difference   between $\alpha=0.5$  and $\alpha=0$. 
and the symbol $\star$ indicates a statistical difference between  
$\alpha^{*}$ and $\alpha=0.5$. Finally, when there is a 
statistical difference between corresponding 
$\gamma=0$ and $\hat{\gamma}>0$, the results of $\hat{\gamma}$ 
are colored in magenta.} 
\label{tab.cm_alpha}
\end{center}
\end{table}

\subsubsection{Effect of adding the initial semantically filtered scores} 
Before a comparison between $rsv^{q_t,vt}_{cm}$ and $rsv^{q_t,tv}_{rw}$, 
let us first compare again the case of  $\gamma=0$ and $\hat{\gamma}>0$, 
but this time
 in the context of Eqs.~\ref{eq:cmtv} and ~\ref{eq:cmvt}. 
The results are shown in 
 Table~\ref{tab.cm_alpha} where we 
varied $\alpha$ between 0 and 1 (with a 0.1 step),
but we  only show the results for $\alpha=0.5$ and $\alpha^*$ 
(which is the value that outperforms all other ones).

From this table we can draw the following conclusions~: 
\begin{itemize}[label=-]
\item Adding $s^{q_t}_t$ to $x^{q_t}_{(1)}$ (Eq. \ref{eq:cmtv}) 
is always a winning strategy whether $\gamma=0$ or $\hat{\gamma}>0$. 
\item Adding $s^{q_t}_v$ to $y^{q_t}_{(1)}$ (Eq. \ref{eq:cmvt}) 
is not always beneficial 
and could be damaging for the search results especially with a
non optimal $\alpha$ value. However, we have to make the distinction 
between the cross-media oriented diffusion process with a prior
($\hat{\gamma}>0$) and the one without a prior ($\gamma=0$).
 In the former case, the graph based scores already benefited from 
$s^{q_t}_v$ (used as a prior). 
As a result, adding the latter semantically filtered visual relevance score to
$y^{q_t}_{(1)}$ by means of a late fusion strategy does not bring any
improvement or worse, it could hurt the performances which is typically 
the case
for $\alpha=0.5$.  On the contrary, when we do not use any prior in the
cross-media oriented diffusion process ($k\leq 30$, $\gamma=0$), then we
always observe a dramatic increase of the MAP measures. 
\item For both Eq. \ref{eq:cmtv} and Eq. \ref{eq:cmvt}, 
$\gamma=0$ is in general  better than $\hat{\gamma}>0$, when $\alpha=0.5$ but
compared to (the best) $\alpha^*$, the results are in 
general very similar (except
for WEBQ) and with no statistical difference between them. 
\item  It is difficult to judge between the two following cases~:
 $\gamma=0$, $\alpha>0$  on the one hand and 
 $\hat{\gamma}>0$ and $\alpha=0$ on the other hand. 
As regard to the cross media diffusion process, the former case 
promotes no prior but a late fusion while the latter case 
rather supports the integration of a prior in the diffusion 
process and no further linear combination. If we compare the latter case, $\hat{\gamma}>0$ and $\alpha=0$, to
the former one but with $\gamma=0$ and $\alpha=0.5$  (no need to tune the $\alpha$ parameter), we see a 
slight advantage of the first setting over the second one.
Therefore in what follows, we will pursue the experiments with the cross-media approach with a prior $\hat{\gamma}>0$ and no late fusion $\alpha=0$.   Nevertheless the other approach can also be 
 considered as a possible multimedia retrieval model. Note that these 
observations  support the fact that the cross-media approach 
gives search results that are complementary to the initial semantically filtered scores and 
this complementarity can be formulated \textit{via} 
an early integration of these scores in the diffusion process 
as priors or \textit{via} a late integration through a linear 
combination with the obtained graph based scores. 
\end{itemize}

Accordingly, when using Eqs. \ref{eq:cmtv} and \ref{eq:cmvt} as multimedia
retrieval models, we recommend to employ by default either 
the parameter setting $\{k=10,\gamma=\hat{\gamma}, \alpha=0\}$ or 
 $\{k=10,\gamma=0,\alpha=0.5\}$. These configurations indeed lead to 
near optimal results on the first three datasets, 
and significantly better MAP values than the ones obtained 
with $s^{q_t}_{v}$ and $s^{q_t}_t$ for all datasets. Concerning, the  
WEBQ the best results we obtained were with $k=l$, $\gamma=0$ and
$\alpha^*$. 


\begin{table}[t]
\begin{center}
 \begin{tabular}{|c|c|c|c|c|}
\hline
   & IAPR  & WIKI10   & WIKI11 & WEBQ\\
\hline
 \begin{tabular}{c}
\\
\end{tabular} &
 \begin{tabular}{ll}
$s^{q_t}_v$ &  $s^{q_t}_t$ \\
\hline
 27.6 & 26.3   \\
\end{tabular} &
 \begin{tabular}{ll}
$s^{q_t}_v$ &  $s^{q_t}_t$ \\
\hline
24 & 26.3   \\
\end{tabular} &
 \begin{tabular}{ll}
$s^{q_t}_v$ &  $s^{q_t}_t$ \\
\hline
18.1 & 27.8  \\
\end{tabular} &
 \begin{tabular}{l}
 $s^{q_t}_t$ \\
\hline
57  \\
\end{tabular}
\\
\hline
 \begin{tabular}{ccc}
 $k$ & $\alpha$ \\
\hline
10 & 0 \\
10 & 0.5 \\
10 & $\alpha^*$ \\
\hline
30 & 0 \\
30 & 0.5 \\
30 & $\alpha^*$ \\
\hline
50 & 0 \\
50 & 0.5 \\
50 & $\alpha^*$ \\
\hline
$l$ &  0 \\
$l$ & 0.5 \\
$l$ & $\alpha^*$ \\
\end{tabular} &
 \begin{tabular}{ll} 
 $rsv_{gd}^{q_t,vt}$ & $rsv_{gd}^{q_t,tv}$   \\
\hline
35.1 & 24.2$^\dagger$ \\
35.1 & \bf{27} \\
\bf{36.1} & \bf{27} \\
\hline 
33.4 & \mk{24.1}$^\dagger$ \\
33.5 & 26.1 \\
34.3 & 26.6 \\
\hline 
\mk{30.7} & \mk{24.6}$^\dagger$ \\
31.6 & 26.2 \\
31.8 & 26.6 \\
\hline 
\mk{31.9} & 26.4 \\
\mk{30.9} & 26.5 \\
\mk{32}$^\star$ &  26.6\\
\end{tabular} &
 \begin{tabular}{ll} 
 $rsv_{gd}^{q_t,vt}$ & $rsv_{gd}^{q_t,tv}$   \\
\hline
28.4 & 29.2 \\
27.5 & 28.7 \\
\bf{28.7}$^\star$ &  29.2 \\
\hline
27.8 & \bf{29.4} \\
27.6 & 28.8 \\
28.2 & \bf{29.4} \\
\hline
\mk{26.2} & \bf{29.4} \\
26.7 & 28.8 \\
\mk{26.8} & \bf{29.4} \\
\hline
25.3 &  \mk{27.9} \\
\mk{25.1} &  \mk{27.5} \\
\mk{25.4} &  \mk{27.9} \\
\end{tabular} &
 \begin{tabular}{ll} 
 $rsv_{gd}^{q_t,vt}$ & $rsv_{gd}^{q_t,tv}$   \\
\hline
22.8 & 31.3\\
22.9 & 30.4 \\
\bf{23.5} & \bf{31.4}$^\star$ \\
\hline
22.1 & 30.9 \\
23 & 30.2 \\
23.2 & 30.9 \\
\hline
20.3 & 29.4 \\
21.6 & 29.4 \\
21.6 & 29.5 \\
\hline
19.1  & 27.6  \\
19.6 & \mk{27.9}\\
19.8 & 28.1 \\
\end{tabular} &
\begin{tabular}{l} 
 $rsv_{gd}^{q_t,tv}$   \\
\hline
67.4$^\dagger$ \\
\mk{66.8}\\
67.4$^\star$ \\
\hline
\mk{69.7}$^\dagger$ \\
\mk{68.9} \\
\mk{69.7}$^\star$ \\
\hline
{\bf \mk{70.4}}$^\dagger$\\
\mk{69.4}\\
{\bf \mk{70.4}}$^\star$\\
\hline
\mk{69.5}$^\dagger$\\
\mk{67.9}\\
\mk{69.5}$^\star$\\
\end{tabular} \\
\hline
\end{tabular}
\caption{{\bf Effect of adding the initial  scores $s^{q_t}_v$ and 
$s^{q_t}_t$ to $gd^{q_t}$ and $rw^{q_t}$.}
We show results of $rsv_{gd}^{q_t,vt}$ and $rsv_{gd}^{q_t,tv}$  
with $\beta=0$,  $i=\infty$ and varying  $k$. Note that $k=l$ correspond to the
$rsv_{rw}^{q_t}$ case.  
The symbol $\dagger$  indicates a statistical difference 
 between $\alpha=0.5$  and $\alpha=0$ and the symbol 
$\star$ indicates a statistical difference between $\alpha^{*}$ and 
 $\alpha=0.5$.  
Finally when there is a statistical difference between the 
corresponding cross-media 
oriented diffusion process ($i=1$ shown in Table~\ref{tab.cm_alpha}) obtained with $\hat{\gamma}$, 
the results are colored in magenta.}
\label{tab.rw_alpha}
\end{center}
\end{table}

In Table~\ref{tab.rw_alpha}, we show the performances of the other graph based
 techniques we are interested in~: the generalized diffusion model and in
 particular, the random walk process (last three rows with $k=l$). 
Hence, comparing Tables~\ref{tab.cm_alpha} and~\ref{tab.rw_alpha} allows us to
 assess Eqs.~\ref{eq:cmtv} and \ref{eq:cmvt} 
against Eqs.~\ref{eq:best_rwtv} and \ref{eq:best_rwvt}. 
In other words, we again compare the two particular cases $i=1$ and $i=\infty$ 
for this new set of experiments. Note that both $rsv_{gd}^{q_t}$ and $rsv_{rw}^{q_t}$
 performed poorly without the prior so for these models, we did not make any experiment 
with  $\gamma=0$ (instead we take $\hat{\gamma}>0$). 

\begin{figure}
\begin{center}
\includegraphics[width=\textwidth]{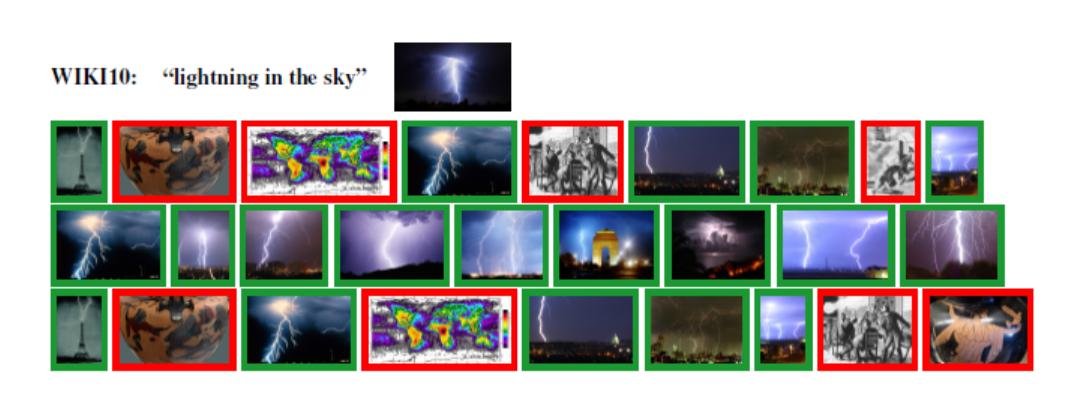}
\end{center}
\caption{Top retrieved images with textual similarity  (second row), with cross-media $rsv_{cm}^{q_t,tv}$
using $k=10$ and $i=1$ (third row) and with random walk   $rsv_{rw}^{q_t,tv}$  (last row),  for the topic 7
at ImageCLEF Wikipedia Challenge 2010 (shown in first row). Green means relevant, red non-relevant.}
\label{fig:light}
\end{figure}

The comparison between the two cases, $rsv_{gd}^{q_t}$ ($k\ll l$) and  
$rsv_{rw}^{q_t}$ ($k=l$), yields to the following observations~:
\begin{itemize}[label=-]
\item As we already pointed out in  Table~\ref{tab.cm_alpha}, adding 
$s^{q_t}_v$ and $s^{q_t}_t$ as priors did not bring any improvement in most cases  and with $\alpha=0.5$ we can 
even observe a decrease in the performances. Hence, when $rsv_{gd}^{q_t}$ 
or $rsv_{rw}^{q_t}$  are used, it is better 
to not recombine these scores again with the semantically filtered monomodal scores but to use directly
$gd^{q_t}$ and $rw^{q_t}$ as multimedia retrieval models.
\item When  $k=10$ and $\gamma=\hat{\gamma}$, 
$rsv_{gd}^{q_t}$ ($i=\infty$) and $rsv_{cm}^{q_t}$ ($i=1$ shown in Table~\ref{tab.cm_alpha}) yield very similar results 
(except for WEBQ), however for larger $k$ values the $rsv_{cm}^{q_t}$ 
outperforms $rsv_{gd}^{q_t}$ 
showing again that for $rsv_{gd}^{q_t}$ it is even more important  to 
use small  $k$  values than for $rsv_{cm}^{q_t}$.
\item For all datasets, $rsv_{gd}^{q_t}$ outperforms $rsv_{rw}^{q_t}$
(except  WEBQ with $k=10$). More interestingly, the results obtained with 
$rsv_{rw}^{q_t}$ remain significantly below the results provided by 
$rsv_{cm}^{q_t}$  on the first three datasets. Concerning  WEBQ,
$rsv_{rw}^{q_t,tv}$ is better than $rsv_{cm}^{q_t,tv}$ with $k\leq 50$, however
for $k=l$ the performances are comparable, especially when we consider
$\gamma=0$ for $rsv_{cm}^{q_t}$. We thus state that cross-media
oriented diffusion processes provide search results that are more
complementary to $s^{q_t}_t$ and $s^{q_t}_v$ than random walk based diffusion
processes since a late fusion (even with equal weighting) gives better results 
in the former case than in the latter case. This is further illustrated 
in Figure \ref{fig:light} on a Wikipedia query. 
\end{itemize}

Finally,  these two latter tables enable us to conclude about the asymmetric 
search scenario. Our experimental results show that, when given a 
text query only, the best multimedia fusion strategies are either to consider 
the cross-media oriented diffusion process
(with or without prior) linearly 
combined with the initial semantically filtered text relevance scores which leads to the retrieval model $rsv_{cm}^{q_t}$
(Eq. \ref{eq:cmtv}),  or to use instead of $rsv_{cm}^{q_t}$, the generalized diffusion model $gd^{q_t}$
without recombination with the initial scores. In both cases, it is important to
use a relatively small $k$ (\eg $k=10$). 

\subsubsection{Combining all relevance scores in a late fusion scheme}
 Finally, we study the combination of both initial relevance 
scores $s^{q_t}_v$ 
and $s^{q_t}_t$ with both  multimedia graph based 
scores $y^{q_t}_{(i)}$ and $x^{q_t}_{(i)}$ in an ultimate 
linear combination. This is possible in the case of the 
symmetric search scenario. We previously presented in section \ref{comp_cm_rw},
Eq. \ref{cm_formula_qt} and Eq. \ref{rw_formula_qt} which refer 
to such combinations. For convenience, we recall these latter formulas below~:
\begin{eqnarray*}
rsv^{q_t}_{cm}(q,.) & = & \alpha_t s^{q_t}_t(q,.)+\alpha_v s^{q_t}_v(q,.)+\alpha_{tv} cm^{q_t}_{tv}(q,.)+ \alpha_{vt} cm^{q_t}_{vt}(q,.)\\
rsv^{q_t}_{gd}(q,.) & = & \alpha_t s^{q_t}_t(q,.)+\alpha_v
s^{q_t}_v(q,.)+\alpha_{tv} gd^{q_t}_{tv}(q,.)+ \alpha_{vt} gd^{q_t}_{vt}(q,.)\\\
rsv^{q_t}_{rw}(q,.) & = & \alpha_t s^{q_t}_t(q,.)+\alpha_v
s^{q_t}_v(q,.)+\alpha_{tv} rw^{q_t}_{tv}(q,.)+ \alpha_{vt} rw^{q_t}_{vt}(q,.)
\end{eqnarray*}
where $cm^{q_t}_{tv}=x^{q_t}_{(1)}$, $cm^{q_t}_{vt}=y^{q_t}_{(1)}$, 
$gd^{q_t}_{tv}=x^{q_t}_{\infty}$, $gd^{q_t}_{vt}=y^{q_t}_{\infty}$ 
with $k \ll l$  and
$rw^{q_t}_{tv}=x^{q_t}_{\infty}$, $rw^{q_t}_{vt}=y^{q_t}_{\infty}$ with $k=l$.\\

According to the results we underlined previously, we chose the following settings~: 
\begin{itemize}
\newcommand{\myitem}{\item[\textbullet]}
\myitem $k=10$ (except for $rw$ where $k=l$)
\myitem $\beta=0$ (no late fusion of similarity matrices).
\myitem $\gamma=0.3$ (using a monomodal prior) or $\gamma=0$ 
(without monomodal prior).
\myitem $\alpha$ are set to uniform weights or to best performing weights.
\end{itemize}
We show the obtained results in Table~\ref{tab.rw_cm_all}.  
In addition, we show the
MAP values obtained by the late fusion of semantically filtered 
relevance scores, $\alpha_v s^{q_t}_v(q,.)+\alpha_t s^{q_t}_t(q,.)$, which represents our baseline.
From this table, we can draw the following conclusions~:
\begin{itemize}[label=-]
\item In a symmetric search setting, combining initial filtered scores with graph based diffusion processes scores is beneficial and performs better than the baseline. The graph based measures are thus complementary to the initial scores.
\item $rsv^{q_t}_{cm}$, $rsv^{q_t}_{gd}$ provide similar performances and outperform the random walk $rsv^{q_t}_{rw}$ method.
\end{itemize}



\begin{table}[!h]
\begin{center}
 \begin{tabular}{|c|c|c|c|c|}
\hline
   & IAPR  & WIKI10   & WIKI11 \\
\hline
$0.5 s_v+ 0.5 s_t$ & 34.5 & 35.2 & 35.4 \\
$\alpha_v^* s_v+ \alpha_t^* s_t$ & 35.4 & 35.2 & 35.4 \\
\hline
$rsv^{q_t}_{cm}, (\gamma=0)$ & \mk{37.3}$^\dagger$ & \mk{35.9}  & 33.6$^\dagger$ \\
$rsv^{q_t*}_{cm}, (\gamma=0)$ & \mk{39.4}&  \bf{\mk{36.1}} &  35.7 \\
\hline
$rsv^{q_t}_{cm},  (\gamma=0.3)$ & \mk{37.3}$^\dagger$ & \bf{\mk{36.1}} &  \bf{36} \\
$rsv^{q_t*}_{cm}, (\gamma=0.3)$ & \bf{\mk{39.5}}&  \bf{\mk{36.1}} &  \bf{36} \\
\hline
$rsv^{q_t}_{gd},  (\gamma=0.3)$ & \mk{37.3}$^\dagger$ & \mk{35.8} &  35.1\\
$rsv^{q_t*}_{gd}, (\gamma=0.3)$ & \mk{38.7}&  \mk{36} &  35.8 \\
\hline
$rsv^{q_t}_{rw},  (\gamma=0.3)$ & 34.4 & 35.4 & \mk{34.2}$^\dagger$ \\
$rsv^{q_t*}_{rw}, (\gamma=0.3)$ & 36.1 & 35.6 &  \mk{35.4} \\
\hline
\end{tabular}
\caption{{\bf Combining all relevance scores in a late fusion scheme.} 
We show the results of $rsv^{q_t}$ with 
uniform weights and  $rsv^{q_t*}$
corresponding to the best linear combinations of the different scores. We considered  $k=10$
(except for $rsv^{q_t}_{rw}$), $\beta=0$ and $\gamma=0.3$. In
addition as we average with the monomodal score,  we also show 
the case of  $\gamma=0$ (no prior)  for the cross-media diffusion process. 
We do not show results with $\gamma=0$ for the random walk and generalized 
diffusion model  as  we have seen that iterating without the prior
yields to  much worse results. The symbol  $\dagger$ indicates a 
statistical difference between uniform
 weights and tuned weights.  We colored in magenta the values of the 
results of $rsv^{q_t}$ and the ones of $rsv^{q_t*}$
when there is a statistical difference with the semantically filtered 
late fusion (with uniform respectively tuned $\alpha$ weights).}
\label{tab.rw_cm_all}
\end{center}
\end{table}

\subsection{Advantage of the cross-media oriented diffusion process}\label{subsec:advantages_cm}

With regard to the comparison between the cross-media and the random walk views
of the proposed unified multimedia retrieval model, the experiments we conducted
favors the former orientation as compared to the latter one. Indeed, all along
this current section we have shown that~: 
\begin{enumerate}[label= (\roman*)]
\item Choosing the probability distribution according to the semantically filtered
  relevance scores $s^{q_t}_t$ and $s^{q_t}_v$ as an initialization of the
  diffusion processes in Eqs. \ref{eq.tv_gen_qt} and \ref{eq.vt_gen_qt}
  respectively is better than the uniform distribution and this also supports
  the transmedia fusion principle in CBMIR.
\item  Using a small neighborhood as
  proxy in the transmedia approach by using $\mathbf{K}(.,k)$ with 
$k\approx 10$ is in general better than choosing a large neighborhood; 
this is even more important when we iterate the generalized diffusion model;
\item  One-step or occasionally two-step walks provide better performances 
than longer walks when we do not use the initial scores as priors.
\item  The cross-media oriented diffusion process without the prior 
is the most complementary to the  initial semantically filtered 
relevance scores and yields to better results than the random walk oriented
 diffusion process with or without the prior. 
\item Adding a prior in the diffusion processes is beneficial but using the
 latter information in a late fusion scheme after having calculated the graph
 based scores using only a single step gives comparable results. Both
 solutions can be considered, however, the $\gamma$ parameter for the 
 former seems to be more stable that the best $\alpha$ value in the 
latter case. Concerning the generalized diffusion model it is important to 
use the  prior and preferably not to combine it with the semantically filtered monomodal scores. 
\item Overall, the best parameter setting we obtained as regard to our
  multimedia retrieval system is a late fusion between semantically filtered
  scores and cross-media oriented diffusion processes scores, the latter using 
the monomodal scores also as prior. 
\end{enumerate}

All these experiments allow us to claim that cross-media diffusion 
processes give better search results than random walk diffusion 
processes both for the asymmetric and symmetric search settings 
as long as the parameters are set correctly and in that perspective,
we have also provided sets of parameters values one should use by default.

Below, we further underline and summarize two other important advantages of the
cross-media framework compared to the random walk and the 
generalized diffusion model from the practical 
standpoint~:

\begin{itemize}[label=-]
\item {\bf Complexity.} 
Our experimental results have shown that cross-media oriented diffusion processes give better results than random walk diffusion processes. As a consequence, we do not need to iterate the diffusion process until convergence and we have shown that a single step was sufficient to reach good performances. Hence, the model underlying the cross-media similarities not only provide better results but it also reduces the computation time. 
  Moreover, when we use the cross-media settings ($\beta=0,k=10,i=1$), we employ
  the nearest neighbor thresholding operator $\mathbf{K}(.,k)$ on the initial
monomodal scores. In that case, we do not even need to compute the $l \times l$
  similarity matrices, but only $k \times l$ similarity values which correspond
  to the rows of $S^{q_t}_t$ and $S^{q_t}_v$ 
associated to the $k$ nearest neighbors (even when using the prior these values
  are to be added simply to the corresponding rows). This further reduces 
the computational cost of the proposed graph based method. Since image 
representation and similarities have much higher time and storage complexities than text, overall, our proposal can easily tackle large collections of multimedia objects.

\item {\bf Score Normalization}
In order to compare the two methods we have been interested in, we normalized
the scores and similarities in order to have probability distributions. This
normalization was necessary for the random walk approach as explained in section
\ref{comp_cm_rw}. However, since we have shown that iterating the diffusion
process more than once was generally damaging, we concluded that we needed to
iterate Eq. \ref{eq.tv_gen_qt} and Eq. \ref{eq.vt_gen_qt} only one time. As a
result, it is no more mandatory to have probability distributions as for
$x^{q_t}_{(i)}$ and $y^{q_t}_{(i)}$. We thus computed $rsv^{q_t}_{cm}$ using
another normalization~: we replaced each score or similarity $s^{q_t}(d,d')$ by
$(s^{q_t}(d,d')-\min\{s^{q_t}(d,.)\})/(\max\{s^{q_t}(d,.)\}-\min\{s^{q_t}(d,.)\})$.
We show in Table \ref{tab.cmrw_normalization}\footnote{Note that for WEBQ, we do
  not have the image query and we can not provide the final scores
  $rsv^{q_t}_{rw}$ and $rsv^{q_t}_{cm}$ given by Eq. \ref{rw_formula_qt} and
  Eq. \ref{cm_formula_qt} unlike for other tasks. In that case, we show the best
  values we obtained among all the previous experimental results we
  presented. Note that we show single step ($i=1$) results in the case of   
$ rsv^{q_t,nm}_{cm}$.}, the performances of the cross-media oriented diffusion process using this other normalization procedure. This is denoted by $rsv^{q_t,nm}_{cm}$. Overall, we reached even better MAP results. As a consequence, this suggests that our study of graph based methods could be further enriched with the impact of other kinds of normalization methods.

\begin{table}[!h]
\begin{center}
 \begin{tabular}{|c|c|c|c|c|c|}
\hline
   & IAPR  & WIKI10   & WIKI11 & WEBQ\\
\hline
$\alpha_v^* s_v+ \alpha_t^* s_t$ & 35.4 & 35.2 & 35.4 & 57 \\
$rsv^{q_t*}_{rw}$     & 36.1  & 35.6  &  35.4  & 69.5 \\
$rsv^{q_t*}_{gd}$     & 38.7  & 36  &  35.8  & 70.4 \\
$rsv^{q_t*}_{cm}$     & 39.5  & 36.1  &  \textbf{36}  & 69.6 \\
$rsv^{q_t,nm}_{cm}$  &   \textbf{40.2} &  \textbf{36.2}    &
35.6    & \textbf{70.7}  \\
\hline
\end{tabular}
\caption{Results with different score normalization. }
\label{tab.cmrw_normalization}
\end{center}
\end{table}

\end{itemize}

\section{Conclusion}\label{discussions}
We have addressed the problem of multimedia information fusion in CBMIR and
compared the cross-media similarities to the random walk methods. First of all,
we have proposed a unifying framework that integrates the text query based
semantic filtering of multimedia scores and similarities and which generalizes
both graph based techniques in a unifying model.  

Furthermore, we have extensively studied many factors that impact the performances of these graph based methods. One of our goals was to provide some guidelines on how to best use those methods for two different multimodal search scenarios : the asymmetric and symmetric cases.
Our findings have been validated on three real-world datasets which are public and accessible to the research community.

All in all, we can summarize our findings about graph based methods as follows :
\begin{itemize}[label=-]
\item The text query based semantic filtering is an efficient fusion method that allows one to restrain the search space to multimedia items that are the most semantically related to the text query. We suggest to apply this first level of fusion before moving forward with graph based methods.
\item Cross-media similarities and random walk based approaches can be
  seamlessly embedded into a unifying framework. The latter general graph based
  method is defined by Eq. \ref{eq.tv_gen_qt} and Eq. \ref{eq.vt_gen_qt} which
  allow one to take into account both the asymmetric and the symmetric search
  scenarios. The unifying framework exhibits transmedia diffusion processes with
  or without priors and bring to light the main differences between the two
  types of methods used by the community. But in a more general scope, it allows
  us to formalize the interesting features and parameters one should pay
  attention to when using an unsupervised  graph based approach in 
content based image/text multimedia retrieval tasks.
\item The experiments we conducted globally show that cross-media oriented
  diffusion processes outperform random walk based methods. Typically, we claim
  that the default setting for Eq. \ref{eq.tv_gen_qt} and Eq. \ref{eq.vt_gen_qt}
  which yields to near best performances on average are :
\begin{itemize}[label=\textbullet]
\item $\beta=0$ (no late fusion between similarity matrices).
\item $\gamma=0.3$ (using a prior helps)
\item $k \approx 10$ (a few nearest neighbors as proxies is better)
\item  $i=1$ (one iteration is sufficient)
\end{itemize}
We have shown that the cross-media oriented 
diffusion process being complementary to the initial 
relevance scores $s^{q_t}_v(q,.)$ respectively
 $s^{q_t}_t(q,.)$ can be successfully 
combined with them to further improve the retrieval accuracy. 
Overall, we obtained the best search results with $rsv^{q_t}_{cm}(q,.)=\alpha_t
s^{q_t}_t(q,.)+\alpha_v s^{q_t}_v(q,.)+\alpha_{tv} cm^{q_t}_{tv}(q,.)+
\alpha_{vt} cm^{q_t}_{vt}(q,.)$ for all real-world tasks we tested especially
when we used another normalization  than the ones required by the 
iterative processes. Last but not least, cross-media oriented 
diffusion processes have a lower computational cost compared to both the
random walk oriented and generalized diffusion  models and hence 
such graph based techniques can tackle large multimedia 
repositories in a scalable way.
\end{itemize}

\appendix

\section{Text representation and similarities}\label{sec:text_models}

Standard pre-processing techniques were first applied to the textual
part of the documents. After stop-word removal, words were lemmatized 
and the collection of documents indexed with
Lemur\footnote{{http://www.lemurproject.org/}}.

We describe here the Lexical Entailment (LE) model used on the Wikipedia dataset as it is a less well-known model.
\cite{berger_sigir99} addressed the problem of IR as a statistical translation problem with the well-known noisy channel model. This model can be viewed as a probabilistic version of the generalized vector space model. The analogy with the noisy channel is the following
one : to generate a query word, a word is first generated from a document and this word then
gets ``corrupted'' into a query word. The key mechanism of this model is the probability
$p(v|u)$ that term $u$ is ``translated'' by term $v$. These probabilities enable us to address a vocabulary mismatch, and also some kinds of semantic enrichments.

Then, the problem lies in the estimation of such probability models. We refer here to a previous work \cite{lexical_entailment_ecir06} on LE models to estimate the probability that one term entails another. It can be understood as a probabilistic term similarity or as a unigram LM associated to a word (rather than to a document or a query). 
Let $u$ be a term in the corpus, then LE models compute a probability distribution over terms $v$ of the corpus denoted by $p(v|u)$. These probabilities can be used in IR models to enrich queries and/or documents and to give a similar effect to the use of a semantic thesaurus. However, LE is purely automatic, as statistical relationships are only extracted once from the considered corpus. In practice, a sparse representation of $p(v|u)$ is adopted, where we restrict $v$ to be one of the $10$ terms that are the closest to $u$ using an information gain metric.

More formally, an entailment or similarity between words, expressed by a conditional probability $p(v|u)$, can be used to rank documents according to the following formula :
\begin{equation}\label{eq:lexical_entailment}
s_t(d_t,d'_t)=p(d_t|d'_t) = \prod_{v \in d_t} \sum_{u} p(v| u) p(u|d'_t).
\end{equation}
where $d_t$ (or $q_t$) and $d'_t$ are two texts, $p(u|v)$ may be obtained by any of the methods described in \cite{lexical_entailment_ecir06} and $p(u|d'_t)$ is the LM of $d'_t$.

Note that this model was essentially rediscovered in \cite{DBLP:conf/sigir/KarimzadehganZ10} and give substantial improvements compared to standard retrieval models (language models, divergence from randomness, information models).
For instance, the LE model obtains a MAP of 26.3\% compared to 22.6\% on the 2010 Wikipedia dataset. Similarly, on the 2011 dataset,
the LE MAP is 27.82\% compared to a 24.3\% an information based model\cite{clinchant_infomodels}.

\section{Image representation and similarities}\label{subsubsec.img_sim}

As for image representations, we used the Fisher
Vector (FV), proposed in \cite{Perronnin07}, an extension of the  popular
 Bag-of-Visual word (BOV) image representation \cite{Sivic03,Csurka04},  
where an image is described by a histogram of quantized local features.
The Fisher Vector, similarly to the BOV, is  based on an 
intermediate representation, the visual vocabulary, which is built on the 
the top of the low-level feature space. In our experiments we used two
types of low-level features, the SIFT-like Orientation Histograms (ORH)
and the local color (RGB) statistics (LCS) proposed in \cite{Clinchant07} 
and built an independent visual vocabulary for both of them.

The visual vocabulary was modeled by a Gaussian Mixture model (GMM)
$p(u|\lambda)=\sum_{i=1}^N w_i {\mathcal N}(u | \mu_i,\Sigma_i)$,
where $\lambda=\{w_i,\mu_i,\Sigma_i;i=1,\ldots,N\}$ is the set of all parameters
of the GMM and each Gaussian corresponds to a visual word. In the case of 
BOV representation, the low-level descriptors $\{u_t;t=1,\ldots,T\}$ of an image
$d_v$, are transformed into a high-level $N$ dimensional descriptor, 
$\gamma(d_v)$, by accumulating over all low-level descriptors and for each Gaussian, the probabilities of generating a descriptor :
\begin{equation}
\label{eq:softBOV}
\gamma(d_v) =  [\sum_{t=1}^{T} \gamma_1(u_t),\sum_{t=1}^{T}
\gamma_2(u_t), \ldots, \sum_{t=1}^{T} \gamma_N(u_t)]
\end{equation}
where
\begin{equation}
\gamma_i(u_t) = \frac{w_i {\mathcal N}(u_t | \mu_i,\Sigma_i)}{\sum_{j=1}^N w_j {\mathcal N}(u_t | \mu_j,\Sigma_j)}.
\end{equation}

The Fisher Vector \cite{Perronnin07} extends this BOV representation  
by going beyond counting measures (0-order statistics) and by encoding 
statistics (up to the
second order) about the distribution of local
descriptors assigned to each visual word. It rather characterizes the low-level features $\{u_t\}_{t=1,\ldots,T}$ of an image $d_v$ by its deviation from the GMM distribution :
\begin{equation}
G_{\lambda}(d_v)  = \frac{1}{T} \sum_{t=1}^{T} \nabla_{\lambda}
\log \left\{\sum_{j=1}^N w_j {\mathcal N}(u_t | \mu_j,\Sigma_j)\right\}
\end{equation}

To compare two images $d_v$ (or $q_v$) and $d'_v$ from two multimedia documents $d$ (or respectively the query $q$) and $d'$, a natural kernel on these gradients is the Fisher Kernel \cite{Perronnin07} :
\begin{equation}
\label{eqn:kernel}
s_v(d_v,d'_v) = {G_{\lambda}(d_v)}^{\top} F_{\lambda}^{-1} G_{\lambda}(d'_v),
\end{equation}
where $F_{\lambda}$ is the  Fisher Information Matrix. As $F_{\lambda}^{-1}$  is symmetric and positive definite, it has a Cholesky decomposition denoted by $L_{\lambda}^{\top} L_{\lambda}$.
Therefore $s_v(d_v,d'_v)$ can be rewritten as a dot-product between normalized
vectors using the mapping $\Gamma_{\lambda}$ with : 
\begin{equation}
\Gamma_{\lambda}(d_v) =  L_{\lambda}\cdot G_{\lambda}(d_v)
\end{equation} which we refer to as the Fisher Vector (FV) of the  image $d_v$.

As suggested in \cite{PSM10}, we further used a square-rooted
and $L2$ normalized versions of the FV and also built a spatial pyramid
\cite{LSP06}. Regarding this latter point, we repeatedly subdivide the  image
into 1, 3 and 4 regions : we consider the FV of the whole image (1x1);
the concatenation of 3 FV extracted for the top, middle and bottom
regions (1x3) and finally, the concatenation of four FV one for each
quadrants (2x2). In other words, the spatial pyramid (SP) we obtained for each image considering both LCS and ORH features 
is given by $8+8=16$ FV. We used the dot product (linear kernel) to compute
the similarity between the concatenation\footnote{Note that we do not
  need to explicitly concatenate all these vectors as
  $\langle [u,v],[u',v']\rangle=  \langle u,u'\rangle + \langle
  v,v'\rangle$.} of all FV for ORH and LCS.


\bibliographystyle{plain}
\bibliography{mir-fusion.bib}

\end{document}